\documentclass{aa}  

\usepackage{graphicx}
\usepackage{txfonts}
\usepackage{xcolor}
\usepackage{hyperref}
\usepackage{array}
\newcolumntype{C}[1]{>{\centering\let\newline\\\arraybackslash\hspace{0pt}}m{#1}}
\newcolumntype{L}[1]{>{\raggedright\let\newline\\\arraybackslash\hspace{0pt}}m{#1}}

\definecolor{dark-red}{rgb}{0.9,0.0,0.0}
\definecolor{dark-blue}{rgb}{0.15,0.15,0.9}
\definecolor{dark-green}{rgb}{0.15,0.8,0.15}
\definecolor{medium-blue}{rgb}{0,0,0.9}
\hypersetup{
    colorlinks, linkcolor=red,
    citecolor={dark-blue} , urlcolor={medium-blue}
}
\newcommand{\cir}{$\mathrm{^{12}C/^{13}C}$\,}
\newcommand{\teff}{$T_{\text{eff}}$\,}
\newcommand{\logg}{$\log$g\,}

\newcommand{\feh}{$[$Fe/H$]$\,}

\newcommand{\mstar}{M$_\star$\,}
\newcommand{\rstar}{R$_\star$\,}
\newcommand{\lstar}{$L_\star$\,}

\begin{document} 
    \title{Evidence of extra-mixing in field giants as traced by lithium and carbon isotope ratio\thanks{Table 2 is only available in electronic form
at the CDS via anonymous ftp to cdsarc.cds.unistra.fr (130.79.128.5) or via https://cdsarc.cds.unistra.fr/cgi-bin/qcat?J/A+A/}}

   \author{Claudia Aguilera-G\'omez
          \inst{1, 3}
          \and
          Mat\'ias I. Jones
          \inst{2}
          \and
          Julio Chanam\'e
          \inst{3}          
          }

   \institute{N\'ucleo de Astronom\'ia, Universidad Diego Portales, Ej\'ercito 441, Santiago, Chile\\
              \email{claudia.aguilera@mail.udp.cl}
         \and
             European Southern Observatory, Alonso de C\'ordova 3107, Vitacura, Casilla, 19001, Santiago, Chile
         \and
            Instituto de Astrof\'isica, Pontificia Universidad Cat\'olica de Chile, Av. Vicu\~na Mackenna 4860, 782-0436 Macul, Santiago, Chile
             }

   \date{}

\titlerunning{Mixing in giants with \cir and Li}

 
  \abstract
   {Although not predicted by standard stellar evolution, the surface abundance of light elements, such as lithium (Li), carbon, and nitrogen, changes during the red giant branch (RGB) as a result of extra-mixing. This is associated usually with thermohaline mixing acting after the RGB bump. Peculiar Li-enriched RGB stars might also be related to either enhanced extra-mixing or pollution from external sources.}
   {We measure the Li abundance and carbon isotopic ratio \cir in a sample of 166 field red giants with $-0.3\leq$[Fe/H]$\leq0.2$, targeted by the EXPRESS radial velocity program to analyze the effects of extra-mixing. }
   {We measure the abundances with spectral synthesis using high-resolution and high signal-to-noise spectra. Multiple-epoch observations needed for exoplanet detection are used to decrease the effects of telluric contamination in \cir measurements.}
   {Due to the prevalence of upper-limits, the Li abundance pattern is complicated to interpret, but the comparison between RGB and core-He burning giants shows effects of extra-mixing consistent with thermohaline. The most Li-enriched giant in the sample, classified as a RGB star close to the RGB bump, has low \cir. Given that the \cir should not be affected by planet engulfment, this does not seem to be the source of the high Li. There is a decreasing correlation between mass and \cir in the RGB and an increasing correlation in the horizontal branch, which, once again, is consistent with thermohaline mixing. Our data also shows a correlation between \cir and [Fe/H]. There is no evident impact of binarity either on Li or on \cir.}
   {Our sample shows behavior that is consistent with additional mixing acting after the RGB bump. 
   The \cir adds new clues to describe extra-mixing, and could well be the best tool to study mixing in red giants. Additional measurements of \cir in field stars would greatly improve our ability to compare with models and understand mixing mechanisms.}


   \maketitle
%

\section{Introduction} \label{sec:intro}

In stars, lithium (Li) is used extensively as a tracer of mixing and stellar processes affecting stellar interiors, given that its abundance has a strong dependency on stellar environmental conditions. Li burns at temperatures and densities that can be found inside low-mass main sequence (MS) stars, near the base of the convective envelope in solar-like stars. Although standard evolutionary models predict that main sequence stars do not change significantly their Li abundance, the large depletion in the Sun and other similar stars suggest that there are other factors to be considered, such as the presence of planets, age, or binarity \citep[e.g.,][]{Baumann2010, Beck2017, Carlos2019}, with different types of non-standard extra-mixing being able to modify the surface abundance of stars \citep[e.g., rotation, diffusion, gravity waves, overshooting][]{Michaud1986, Pinsonneault1992, CharbonnelTalon2005}, and create features such as the Li dip \citep{BoesgaardTripicco1986, DelgadoMena2015} and the Li desert \citep{Ramirez2012, AG2018}.

Once low-mass stars evolve into the red giant branch (RGB) phase, the convective envelope deepens in mass, mixing the Li content of the envelope with the processed, Li-depleted interior, decreasing the surface Li abundance of the star \citep{Iben1965}. This process, the first dredge-up, can also change the surface abundances of other elements, such as carbon and nitrogen \citep{Gratton2000}. Additionally, the carbon isotope ratio \cir is also very sensitive to mixing and evolutionary phase \citep{Gilroy1989,Charbonnel1995}. The photospheric Solar value of \cir$\sim87$ \citep{Scott2006} decreases by a factor of 3 during the first dredge-up \citep{Sweigart1989}, reaching values of about \cir$\sim25$ \citep{DenissenkovHerwig2004}, numbers based on both model predictions and observations of red giants.
After the RGB luminosity function bump, i.e., when the H-burning shell erases the discontinuity in mean molecular weight left by the convective envelope during the first dredge-up, several abundances changes are observed, including a decrease in carbon isotopic ratio, which can get to $\sim10$ or lower values \citep{CharbonnelLagarde2010, Tautvaisien2016}, and Li \citep[e.g.][]{Lind2009}. This new set of abundance changes is thought to be produced by a combination of mixing mechanisms, such as rotationally-induced and thermohaline mixing \citep{Lagarde2019,Charbonnel2020}, with rotational mixing developing even before this point in stellar evolution, but with thermohaline mixing only being able to act after the RGB bump. When the star reaches the tip of the RGB it undergoes the He-flash. Although its effects on stellar abundances are not clear, this dramatic event, happening for stars with masses $\mathrm{M} \lesssim 2.2\ \mathrm{M_{\odot}}$ \citep[although there is a dependency on metallicity;][]{Karakas2010}, alters the structure and could change the abundances of the giant as it enters the horizontal branch (HB) phase.

The internal transport processes affecting the surface abundances after the luminosity bump seem to be strongly metallicity-dependent \citep{Martell2008,Shetrone2019} also depending on stellar mass \citep{Magrini2021}. Thermohaline-mixing is currently the mechanism associated to these changes of chemical abundances after the bump \citep[e.g.][]{Charbonnel2020}. However, it has been noted that for intermediate mass stars  $1.7 \lesssim \mathrm{M}/\mathrm{M_{\odot}\lesssim 2.2}$, rotationally-induced mixing plays a role just as important as thermohaline mixing \citep{CharbonnelLagarde2010}. Higher mass stars should not go through thermohaline mixing since the hydrogen-burning shell never reaches the discontinuity left by the first dredge-up before the star evolves off the RGB. Thus, no RGB bump is present in these stars.

Additionally, a small percentage of giants are enriched in Li. The puzzle of Li-rich giants has existed for several decades, since \citet{WallersteinSneden1982} found the first Li-enriched giant. Some giants have Li abundance values that even exceed the meteoritic $A$(Li)\footnote{$\mathrm{A(Li)}=\log(n_{Li}/n_{H})+12$}=$3.3$ \citep{LoddersFegley1998}, representative of the Solar System abundance at its formation. These are the super Li-rich giants \citep{Monaco2014,Yan2018,Singh2021}. The Li-enriched giants are found in different parts of our Galaxy \citep[][just to name a few]{Gonzalez2009, Monaco2011, Molaro2020}, open clusters \citep{AT2013,Magrini2021}, and globular clusters \citep{Kirby2016, AG2021}. These giants seem to be particularly rare in globular clusters.

Two subsets of solutions have been proposed to this Li-rich giants puzzle. The first set of solutions relies on internal production of Li through Cameron-Fowler mechanism \citep{CameronFowler1971} and posterior rapid extra-mixing of that Li or Beryllium (Be) to outer layers of the star, where the Li is preserved and can be observed. If it is not fast enough, a mixing mechanism may end up decreasing even further the Li abundance, such as the effect observed after the luminosity function bump. The second set of solutions considers pollution of the giant from an external source. Either binaries \citep{Zhang2020,Goncalves2020} or sub-stellar mass companions \citep{Alexander1967,AG2016,SoaresFurtado2021} have been suggested as possible sources of Li. It is possible that the presence of a binary companion is relevant, not because of pollution by direct mass transfer, but because tidal spin-up driven by a binary companion could generate internal Li production \citep{Casey2019}.

The exact evolutionary stage of Li-enriched giants is extremely relevant to identify the mechanism behind the enrichment given that the physical processes that can increase the Li of stars happen predominantly on a specific part of the HR diagram. The recent use of large spectroscopic samples of stars, e.g., Gaia-ESO \citep{Magrini2021all}, LAMOST \citep{Gao2019,Singh2021}, and GALAH \citep{DeepakReddy2019, Deepak2020}, sometimes coupled with asteroseismic information from Kepler and K2 \citep[e.g.][]{Singh2019, Yan2021} or CoRoT \citep[e.g.][]{DeepakLambert2021}, has shown that a large number of enriched giants are located in the core Helium burning phase instead of the first ascending RGB \citep{Yan2021}. This could point toward a physical mechanism acting on the He-flash or upper RGB \citep{Kumar2020}. 

Although the most enriched giants seem to be located in the core-He burning phase, not all of them are, and, although it has been discussed if all these stars go through a process of extra-mixing to increase their Li, it does not seem to be a requirement to understand the Li distribution of the entire population of core-He burning giants \citep{Magrini2021,Chaname2021}.
There are differences in the distribution of Li abundances and fraction of enriched giants in these two evolutionary phases \citep{Yan2021, Martell2021}, something that could indicate that different mechanisms of Li-production are acting. If a mechanism of enrichment on the He-flash \citep{Schwab2020,Mori2021} can explain Li-enriched clump giants, the problem of Li-rich RGB stars still remains.

The planet engulfment process, which could be related to some of the Li-enriched RGB stars \citep{SiessLivio1999}, is interesting in itself, as it can lead to insights on planet-star interactions and planetary system properties even around more massive stars. However, theoretical models \citep{AG2016} show that, under optimistic conditions, Li can increase in the giant after engulfment up to $A$(Li)=$2.2$ dex, and a hot Jupiter-like planet may not even produce a significant signature \citep{SoaresFurtado2021}. As such, this process alone cannot explain the full sample of known Li-rich giants, something that is expected given the preferred location of the most enriched giants on the HB. 
Other signatures in the atmosphere of giants allow to distinguish between enrichment mechanisms. Fast rotation \citep{Privitera2016} and infrared excess \citep{Rebull2015,Mallick2022} can be related to a planet engulfment event, as well as an increase in the Li-6 abundance \citep{AG2020}. The \cir should not change (or could mildly increase) with planet accretion. \citet{Carlberg2012} studied the correlation of \cir with rapid rotation and enhanced Li. Although this was not done to find particular accretion events, the study of global trends in different populations could be evidence of accretion events happening overall. The chrosmopheric activity has also been used as a possible signature to identify the different Li-enrichment processes \citep{Goncalves2020, Sneden2022}.

Regardless of how important the engulfment process is for the Li enrichment, it is expected to happen as the stars evolve, and characterizing its signatures is still needed. On the other hand, if a mixing mechanism is invoked to produce the high Li abundance in some RGB stars, it is necessary to understand if it affects all RGB stars, only a subset depending on stellar parameters, and how this abundance decreases with stellar evolution. A solution to the Li-enriched RGB stars is not clear, but a better characterization of the normal abundance pattern of giants as a function of stellar mass and metallicity, a proper characterization of the mixing mechanisms acting on the RGB after the bump, and a clear way to distinguish signatures of each enrichment process could help in finding a solution. To understand the Li abundance pattern in evolved stars, we must consider that Li-enriched red giant branch star should also experience the mechanism of Li depletion that acts on every other normal giant and depletes their Li abundance (i.e., producing the decrease after the luminosity function bump).

In this work, we analyze the Li abundance and the \cir of a sample giants that have been monitored since 2009 by the EXoPlanets aRound Evolved StarS (EXPRESS; \citealt{Jones2011}) radial velocity program. In Section \ref{sec:obs} we describe the sample and observations. We summarize the procedure to measure stellar parameters in Section \ref{sec:params}, as well as our determination of $A$(Li) and \cir. We describe the general trends between parameters and abundances in Section \ref{sec:results}, discussing the presence of extra-mixing in Section \ref{sec:mixing},  binarity in Section \ref{sec:binary}, and present our summary in section \ref{sec:summary}.

\section{Stellar sample and observations} \label{sec:obs}

The sample of stars analyzed here corresponds to that targeted by the EXPRESS radial velocity (RV) program. This sample is comprised of a total of 166 relatively bright (V $\le$ 8.0 mag) evolved stars, which are observable from the southern hemisphere ($\delta$ $\le$ +20 deg). The stars were selected according to their position in the HR diagram (0.8 $\le$ B-V $\le$ 1.2; -4.0 $\le$ M$_V$ $\le$ 0.5), distance (D $\le$ 200 pc) and photometric stability (H$_p \le 0.015$ mag), thus allowing to include first ascending RGBs and clump stars. Further details of the selection criteria can be found in \citet{Jones2011} and \citet{Soto2021}. \newline

Each of the stars analyzed here was observed by the EXPRESS RV program between 2009 and 2022, using FEROS \citep{feros}, CHIRON \citep{chiron}, HARPS \citep{harps}, and FIDEOS \citep{FIDEOS}, being all of them, high-resolution echelle spectrographs optimized for precision RVs.  
Table \ref{table:spectrographs} lists the main specifications of these four instruments. On average, we have collected $\sim$ 15 individual spectra for each star at different epochs.
We note that HARPS does not cover the wavelength region around 8000 \AA\, used for the \cir analysis (see section \ref{sec:mC12C13}) and therefore it was used only to measure the lithium abundance. For the observations, we adopted different exposure times depending on the stellar magnitude, so that a SNR $\gtrsim$ 100 per resolution element is reached for each spectrum. The typical SNR of single spectra for all four different instruments are also listed in Table \ref{table:spectrographs}. In addition, we combined all of the individual spectra to compute a very high SNR template for each star, as explained in \citet{Jones2017}. In the case of FEROS data, these templates were used to compute the atmospheric stellar parameters (see \citealt{Soto2021}), while for CHIRON and HARPS data we simply used the highest SNR observed spectrum.

\begin{table}
\caption{Instrumental spectral resolution ($\mathrm{\lambda/\Delta \lambda}$), wavelength coverage ($\Delta \lambda$) and typical SNR per pixel of individual spectra used in this work.}              
\label{table:spectrographs}
\centering                                      
\begin{tabular}{c c c c c}          
\hline\hline                        
 &FEROS & HARPS & CHIRON & FIDEOS\\    
 \hline
Resolution & 48000 & 120000 & 79000 & 43000 \\
$\Delta \lambda$ [nm] & 350-920 & 380-690 & 415–880 & 420-800\\
SNR & 150-220 & 120-180 & 140-200 & 100-150   \\
\hline                                             
\end{tabular}
\end{table}

\section{Stellar parameters and abundances} \label{sec:params}

\subsection{Atmospheric and physical parameters \label{sec:atmospheric_pars}}

The atmospheric (\teff, \logg, \feh and microturbulence velocity) and physical stellar parameters (\mstar, \rstar, \lstar and age) were derived using the latest version of the Spectroscopic Parameters and atmosphEric ChemIstriEs of Stars (SPECIES;  \citealt{species2018}; \citealt{Soto2021}) code. Briefly, SPECIES uses an automatic version of the MOOG code \citep[2017 version,][]{sneden1973}, which compares the equivalent widths (EWs) of a list of Fe\,{\sc i} and Fe\,{\sc ii} lines to those derived by solving the radiative transfer equation under the assumption of local thermodynamical equilibrium. 
The derived parameters correspond to an improved version of the results presented in \citet{Jones2011}. The main improvements with respect to the original catalog are: 1) higher quality spectra (which are used to compute the EWs), 2) a refined iron line list, 3) more accurate parallaxes from the Gaia mission \citep{GaiaDR2}, 4) new extinction maps \citep{bovy2016} and 5) a different sets of evolutionary tracks from the MESA Isochrones and Stellar Tracks \citep[MIST;][]{Dotter2016}, which include the equivalent evolutionary points (EEPs) in the interpolation of the isochrone grid to obtain a more precise age and evolutionary status of each star. The stellar atmospheric and physical parameters of the whole sample are listed in Table \ref{table:1}.

\subsection{Stellar abundances: lithium}

We derive abundances for a total of 166 giants. Li abundances are measured by using spectral synthesis around the Li doublet at $\sim$\,6708\,\AA \,to account for blends with other nearby elements. We produce the synthetic spectrum in a range of $10$\ \AA\ around the line using the 2017 version of MOOG, with the ATLAS9 \citep{ATLAS9} atmospheric models. We use the molecular and atomic line list of \citet{Melendez2012}.

\begin{figure}
\centering
 \includegraphics[width=0.5\textwidth]{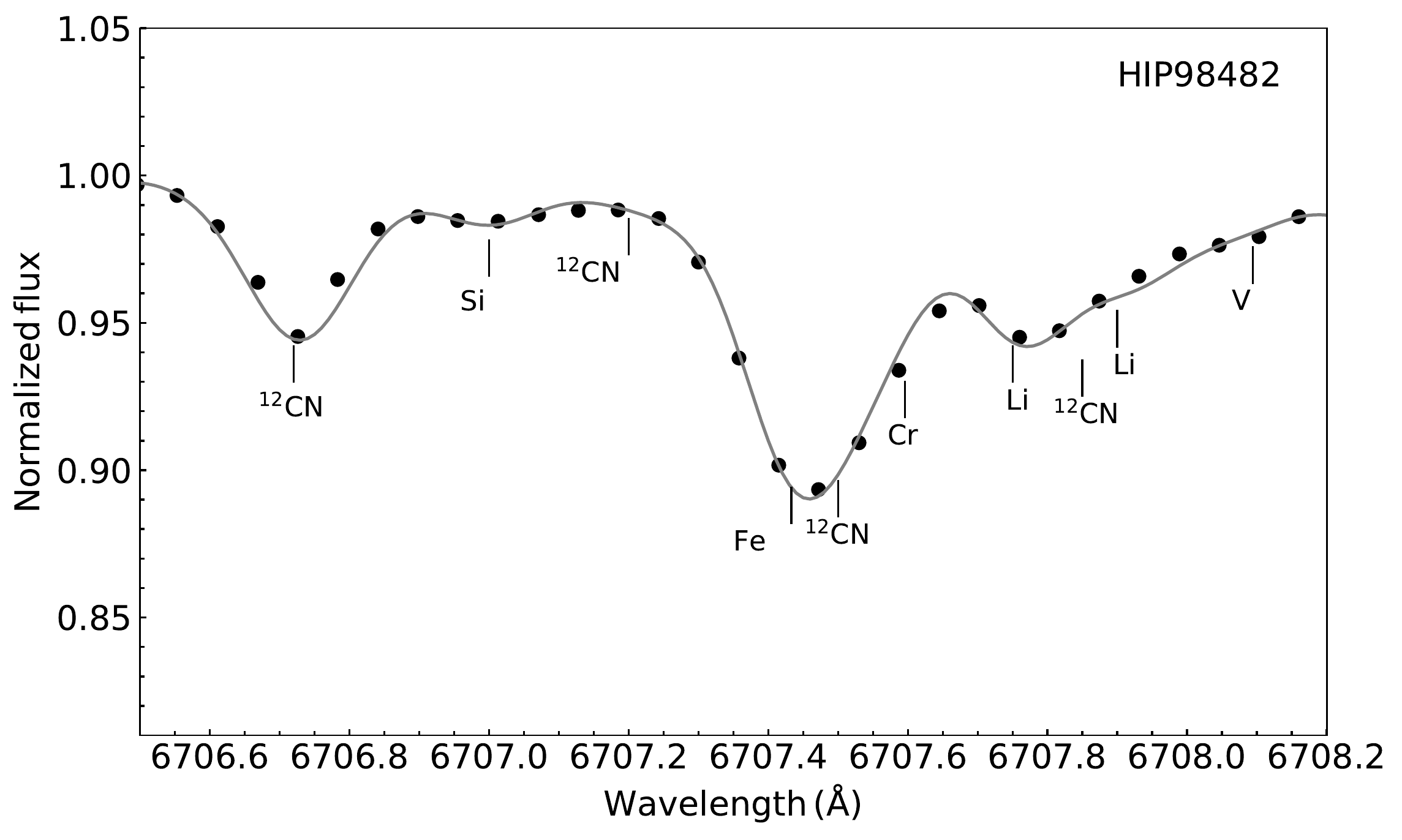}
  \includegraphics[width=0.5\textwidth]{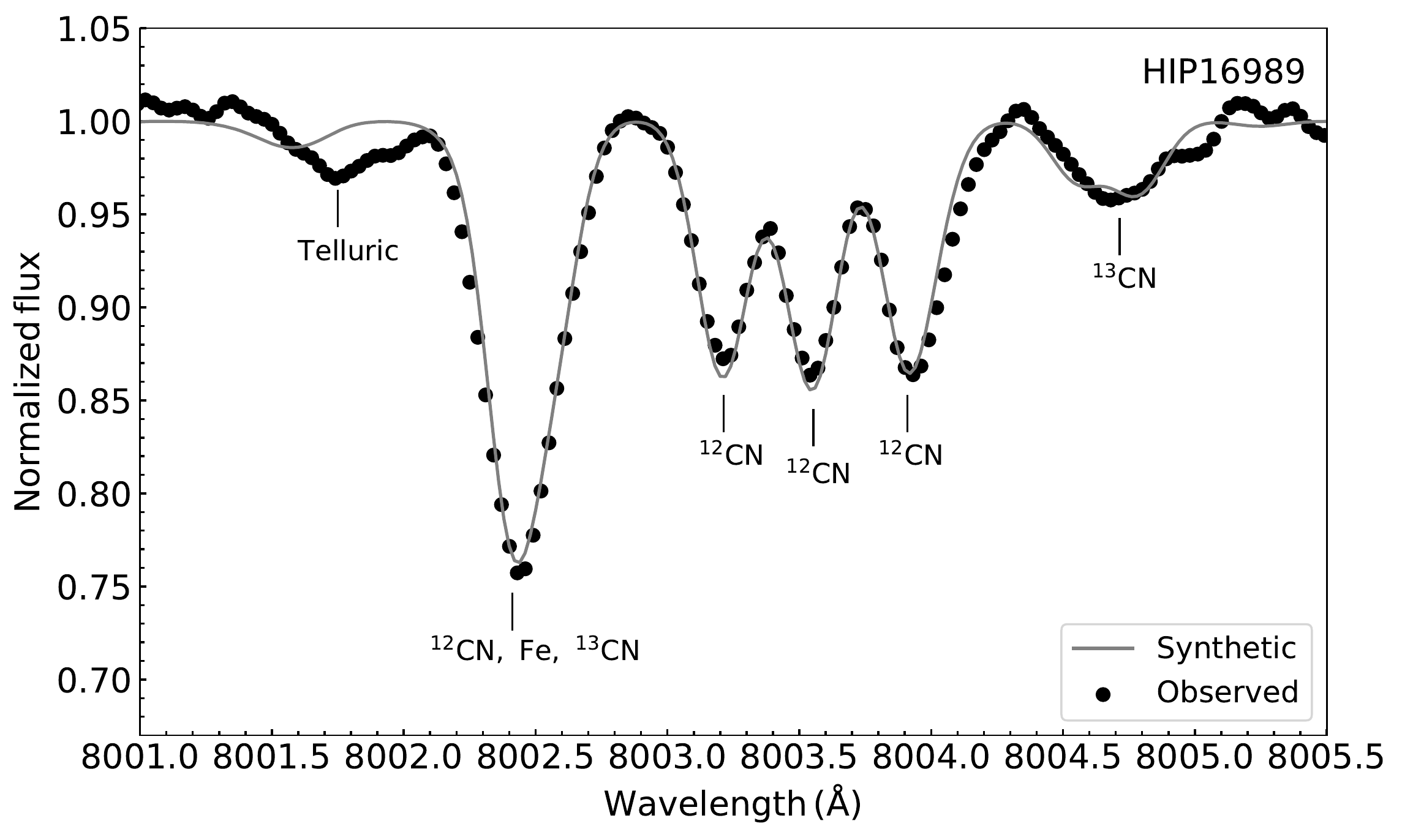}
      \caption{Example fits of synthetic spectra to observed FEROS spectra to measure the Li abundance (top) and \cir (bottom). A label for each important feature is included.}
    \label{fig:example}
 \end{figure}

\begin{figure*}
\centering
 \includegraphics[width=\textwidth]{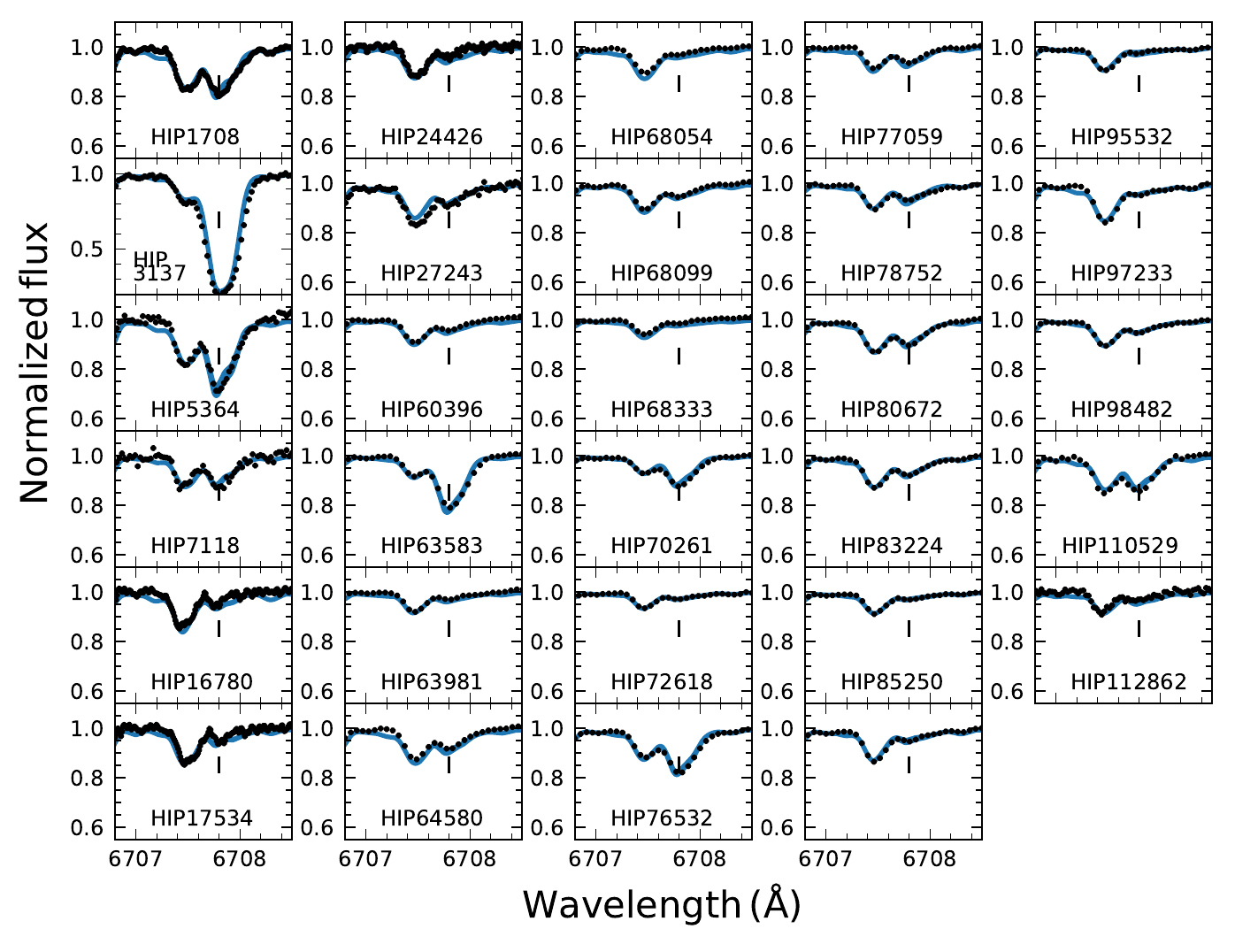}
 
      \caption{Fit of synthetic spectra (blue) to observations (black points) around the Li line, marked with a small vertical line in each panel, for the 29 detections in our sample. Notice that the normalized flux range for the strong Li line in HIP3137 is different from all other fits.}
    \label{fig:detection}
 \end{figure*}

The Li abundances are then obtained by minimizing the difference between the produced synthetic spectra and observations. Small changes to the continuum position, the radial velocity, and the spectral broadening are allowed to produce the best fit. Figure \ref{fig:example}, top panel, shows an example fit of the synthetic to observed spectrum in the region around the Li doublet and typical features in that region. We were able to detect Li in 28 of the 166 giants in our sample. All the remaining stars only have Li upper limits. Figure \ref{fig:detection} shows the synthetic and observed spectrum for each of these detections.

The error in Li abundance considers the uncertainty in continuum positioning, which is added in quadrature to the uncertainty in Li obtained by propagation of errors in stellar parameters. The largest errors due to atmospheric parameters are produced by the effective temperature, with changes in $\log$ g and [Fe/H] producing only negligible variations in Li abundance. The uncertainties in \teff translate into differences in Li abundance of $\mathrm{\Delta A(Li)}\sim0.10-0.15$ dex. The errors in fitting procedure and position of the continuum depend on the quality of the spectra, and are comparable to those produced by the differences in effective temperature. Typical uncertainties for the Li abundance of the sample are $\mathrm{\Delta A(Li)}=0.15$ dex.

We calculated non local thermodynamic equilibrium corrections using the work of \citet{Lind2009NLTE}, through the INSPECT database\footnote{www.inspect-stars.com}. Corrections can be important for giants in our sample, with a mean difference of $0.19$ dex, and a maximum of $0.27$ dex. A few stars are right outside the limits of the grid. For these we use the closest grid point for the corrections. The measured Li abundances or upper limits for all 166 stars are listed in Table \ref{table:1}.

\subsection{Stellar abundances: Carbon isotope ratio \label{sec:mC12C13}}

To measure the carbon isotope ratio \cir we follow a similar procedure to that described in \citet{Carlberg2012}.
We measure the carbon isotope ratio by spectral fitting of CN bands in the region between $8000$ and $8006$ \AA\ using MOOG, with the linelist provided by \citet{Carlberg2012}. We keep a fixed carbon to nitrogen ratio but allow both elements to change from their solar abundances. 
The process to find the carbon isotopic ratio consists in finding the best fit for the $\mathrm{^{12}CN}$ bands between $8002$ and $8004$ \AA\ first, fixing the carbon abundance. After that, we obtain the $\mathrm{^{12}C/^{13}C}$ by fitting the $\mathrm{^{13}CN}$ line at $\sim 8005$ \AA. See Figure \ref{fig:example}, bottom panel, for a sample fit of the \cir region, including common features and telluric lines.

The measurement of carbon isotopic ratio in the optical region is sometimes complicated by the presence of telluric lines. One of the main advantages of analyzing giants targeted by the EXPRESS program is the large amount of individual spectra available for each star at different epochs (typically $\sim$ 15\,-\,20 spectra per star). To identify which spectra are contaminated by telluric lines, we use the atlas telluric spectrum of \citet{Hinkle2000} and compare the position of these features with each of the available spectra for the sample. We discard individual spectra where the $\mathrm{^{13}CN}$ band is contaminated and calculate the isotopic ratio for the remaining spectra.

Additionally, we allow small changes to radial velocity, broadening, and continuum placement to produce the best possible fit between our synthetic spectra and data. One of the largest sources of uncertainty in the fitting procedure is the continuum placement that can be complicated by the presence of multiple lines in the region.

For each star, we adopted the median \cir value from different spectra as the final value. The corresponding 1-$\sigma$ uncertainty is derived from the standard deviation of all single measurements. 
For those stars where only one or two non-contaminated spectra are available we simply adopt a value corresponding to the mean uncertainties for stars with an equal \cir within \cir$\mathrm{\pm 10}$. In general, uncertainties are larger for larger \cir values. A large difference in the ratio only produces a very slight change in the $\mathrm{^{13}C/N}$ spectral region when \cir$\sim50$. Notice that, because of the low sensitivity of the feature for higher \cir values, \cir$=50$ is the higher value in our grid. For \cir$<20$ the typical errors are $\Delta$\cir$=1-3$, while for higher \cir, uncertainties are $\Delta$\cir$=5-8$. 
We were able to measure the \cir only for 134 stars, since for 32 stars either the available spectra does not cover the wavelength region around $\sim$ 8000 \AA\, (e.g. for HARPS data), or because all of the available spectra are strongly contaminated by telluric features in that wavelength region. The derived \cir for the 134 giants are listed in Table \ref{table:1}, as well as the standard deviation and number of spectra used to measure the \cir value.

\subsection{Comparison with other works}

We check our Li measurements by comparing with other catalogs that have computed stellar parameters and Li abundances. We have 17 stars in common with \citet{Liu2014}, 26 giants in common with \citet{Luck2015}, and 41 stars in our work also were analyzed by \citet{Charbonnel2020}. We also have 12 stars in common with \citet{LuckHeiter2007}, which are also included in the updated version of the catalog \citep{Luck2015}. We also find 10 of our giants in the AMBRE catalog with stellar parameters \citep{DePascale2014, Worley2016}, with 7 of them having reported Li abundances \citep{Guiglion2016}. Finally, we have one target in common 
with \citet{Lebre2006}. In total, there are 65 different stars that already have measured Li abundances between these catalogs. We also note that there are no stars in common with the GALAH Survey.

\begin{figure}
\centering
 \includegraphics[width=0.5\textwidth]{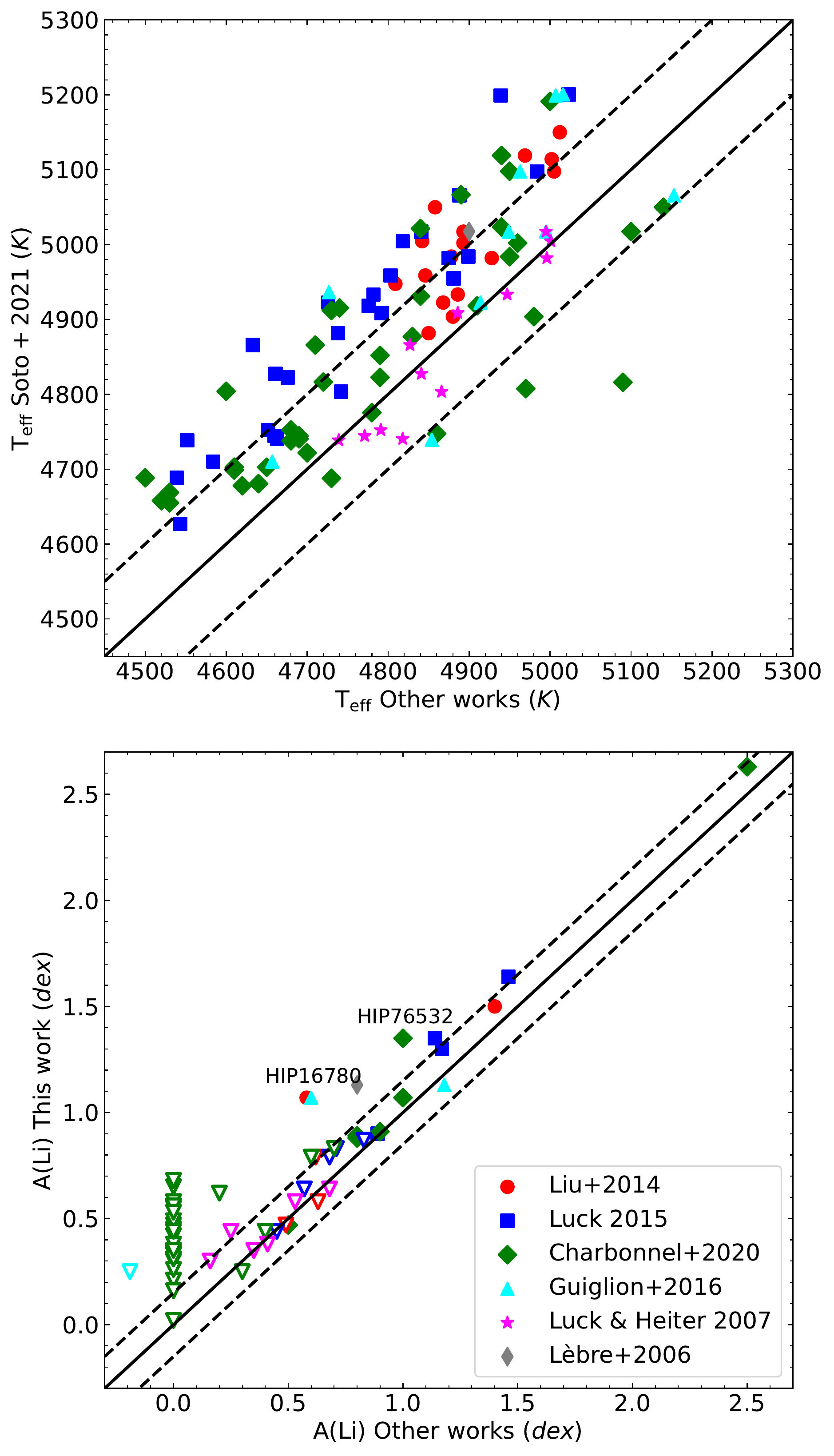}
      \caption{Top: Comparisons in effective temperature between values used in this work to calculate abundances \citep{Soto2021} and the catalogs of \citet{Liu2014} (red circles), \citet{Luck2015} (blue squares), \citet{Guiglion2016} (cyan triangles), and \citet{Charbonnel2020} (green diamonds). Bottom: Comparison between measured Li abundances and reported values in those same studies. We include our upper limits (downward triangles) but remove upper limits from the literature. Dashed lines represent differences of $100$ K in effective temperature and $0.15$ dex in $A$(Li) from the 1 to 1 relation.}
    \label{fig:comparison}
 \end{figure}

Figure \ref{fig:comparison} shows the difference in effective temperatures between literature references and the reported value in \citet{Soto2021}, used in this work (top panel). This is included given the significant impact of this parameter on the Li abundance determination. The bottom panel of the figure compares our measured LTE $A$(Li) with abundances from other works.

In general, effective temperatures from \citet{Soto2021} tend to be higher, with a general offset of $\sim100$ K, and differences that can be as large as $200$ K. In contrast, when comparing with \citet{LuckHeiter2007}, there is no evident offset and the difference between effective temperatures is within the uncertainties. There is a difference of $\sim 100$ K between effective temperatures, similar to the general trend found in comparison with other works. 
Notice that the temperatures in other studies are measured using different methods, such as photometry or fitting synthetic spectra, and thus it is not unexpected to find discrepancies. However, this could impact the determination of Li (with a $\sim100$ K difference in temperature implying $\mathrm{\Delta A(Li)}\sim0.15$), so it is important to keep in mind the effect of parameters on measured abundances.

We only compare our Li abundances with measurements in other works, and no upper limits. The value of upper limits can change depending on the quality of the spectra, thus directly comparing these values would not necessarily provide information about the sample. We do include our upper limits, to make sure they have Li above the determinations in other works to be fully consistent. In spite of the difference between effective temperatures in these works, most of the Li abundances are within a $0.1$ dex difference. There are 4 stars (no upper limits) that show a larger $A$(Li) discrepancy.

In the case of HIP16780, it has a measured Li abundance both in \citet{Liu2014} and \citet{Guiglion2016}, while \citet{Luck2015} report an upper limit. We find a much higher abundance than these 2 other works with measurements, which cannot be explained by the difference in any of the adopted stellar parameters. In fact, by measuring the Li abundance using the atmospheric parameters reported by the AMBRE project we get a value of $A$(Li)=$0.84$ dex, still significantly above their reported abundance of $A$(Li)$=0.60$ dex. 
The upper limit in \citet{Luck2015}, $A$(Li)$\le0.69$ dex is also consistent with measurements in other works.  We notice that the feature we find is small (see Figure \ref{fig:detection}), but still above the noise level. Thus, it is not clear what is producing this large difference between our value and that reported in the literature. Similarly, for HIP76532 we measure a LTE abundance of $A$(Li)=$1.35$ dex, while \citet{Charbonnel2020} reports a value of $A$(Li)=$1.00$ dex. The $30$ K difference in the adopted temperature for the fit cannot explain the offset, considering that a variation of $100$ K could produce a change of at most $0.2$ dex in $A$(Li). The inconsistent metallicity between works cannot explain the change in $A$(Li) either. For that same star, \citet{Luck2015} report an abundance of $A$(Li)$=1.14$ dex.

\section{Results} \label{sec:results}

\begin{sidewaystable*}
\caption{Catalog of parameters and measured Li and \cir\ for the sample giants. Measurements of \cir were possible for 134 stars.Uncertainties in mass are asymmetrical, thus 2 values are included ($M^{+e_\mathrm{Mass}1}_{e_\mathrm{Mass}2}$). E.S. is the evolutionary stage of the giant, either HB or RGB, with $\mathrm{P_{RGB}}$ being the probability of the star being an RGB. The $\mathrm{Li_{flag}}$ entry is 0 for upper limits and 1 for measurements, and $A$(Li)$_{N}$ is the NLTE abundance. $\sigma_C$ is the standard deviation of our \cir measurement while $N_c$ is the number of spectra used to measure the \cir. Only a portion of the table is shown here as a reference. The full catalog is available online.}             
\label{table:1}
\centering
\begin{tabular}{lcccccccccccccC{0.45cm}ccC{0.65cm}C{0.5cm}l}
\hline\hline                 
Name & \teff & $e_{T_{\text{eff}}}$ & \logg & {\tiny $e_{\log g} $} & \feh & {\tiny $e_\mathrm{[Fe/H]}$} & Mass & $e_\mathrm{Mass}$1& $e_\mathrm{Mass}$2 & Age & $\log L$ & $\mathrm{P_{RGB}}$ & E.S. & Flag & $A$(Li) & {\footnotesize $A$(Li)$_{N}$ }& {\footnotesize \cir} & $\sigma_C$ & $N_c$\\ 
- & (K) & (K) & - & - & (dex) & (dex) & ($\mathrm{M_\odot}$) & ($\mathrm{M_\odot}$) & ($\mathrm{M_\odot}$) & (Gyr) & ($\mathrm{L_\odot}$) & - & - & Li & (dex) & (dex) & - & - & - \\ 
\hline                        
HIP242 & 4950.4 & 50.0 & 3.128 & 0.087 & -0.05 & 0.054 & 1.40 & 0.089 & 0.087 & 3.07 & 1.12 & 1.000 & RGB & 0 & 0.40 & 0.566 & 41.3 & 4.2 & 7  \\
HIP343 & 4740.5 & 50.9 & 2.441 & 0.117 & -0.03 & 0.055 & 1.38 & 0.150 & 0.140 & 3.26 & 1.76 & 0.018 & HB & 0 & 0.35 & 0.581 &   &   & 0   \\
HIP655 & 4710.2 & 51.6 & 2.357 & 0.110 & -0.05 & 0.058 & 1.27 & 0.163 & 0.179 & 4.24 & 1.76 & 0.021 & HB & 0 & 0.30 & 0.540 &   &   & 0   \\
HIP671 & 4881.6 & 50.0 & 2.629 & 0.103 & -0.13 & 0.056 & 1.80 & 0.145 & 0.472 & 1.44 & 1.76 & 0.015 & HB & 0 & 0.50 & 0.686 &   &   & 0  \\
HIP873 & 4903.7 & 50.0 & 2.908 & 0.073 & 0.02 & 0.056 & 1.68 & 0.112 & 0.097 & 1.76 & 1.41 & 1.000 & RGB & 0 & 0.58 & 0.767 &   &   & 0  \\
HIP1230 & 4866.7 & 51.1 & 2.866 & 0.114 & -0.13 & 0.054 & 1.31 & 0.096 & 0.097 & 3.79 & 1.34 & 1.000 & RGB & 0 & 0.46 & 0.640 & 31.0 & 4.0 & 7 \\
HIP1684 & 4873.0 & 66.8 & 3.293 & 0.135 & 0.19 & 0.056 & 1.62 & 0.087 & 0.088 & 2.17 & 1.09 & 1.000 & RGB & 0 & 0.60 & 0.805 & 39.7 & 2.7 & 6  \\
HIP1708 & 5065.7 & 50.0 & 2.765 & 0.106 & 0.09 & 0.055 & 2.79 & 0.075 & 0.089 & 0.49 & 1.98 & 0.020 & HB & 1 & 1.64 & 1.800 &   &   & 0  \\
HIP3137 & 4702.7 & 62.6 & 2.377 & 0.131 & -0.04 & 0.059 & 1.61 & 0.120 & 0.129 & 1.99 & 1.65 & 0.922 & RGB & 1 & 2.63 & 2.652 & 13.7 & 1.2 & 0 \\
HIP3436 & 4758.6 & 63.5 & 2.758 & 0.128 & 0.03 & 0.057 & 1.14 & 0.111 & 0.097 & 6.86 & 1.21 & 1.000 & RGB & 0 & 0.32 & 0.541 & 41.7 & 3.0 & 7  \\
HIP4293 & 4737.9 & 50.0 & 2.694 & 0.092 & -0.07 & 0.057 & 1.50 & 0.115 & 0.120 & 2.47 & 1.57 & 1.000 & RGB & 0 & 0.26 & 0.478 &   &   & 0 \\
HIP4587 & 5017.2 & 50.0 & 2.691 & 0.082 & -0.19 & 0.052 & 2.07 & 0.159 & 0.206 & 1.11 & 1.80 & 0.005 & HB & 0 & 0.58 & 0.738 &   &   & 0 \\
HIP4618 & 4703.2 & 55.3 & 2.723 & 0.114 & -0.11 & 0.055 & 0.97 & 0.054 & 0.036 & 11.37 & 1.24 & 1.000 & RGB & 0 & 0.00 & 0.220 & 47.7 & 0.0 & 10 \\
HIP5364 & 4626.9 & 64.0 & 2.402 & 0.139 & -0.05 & 0.057 & 1.43 & 0.129 & 0.134 & 2.89 & 1.80 & 0.018 & HB & 1 & 1.30 & 1.543 & 19.0 & 0.0 & 1  \\
HIP6116 & 4775.4 & 50.9 & 3.028 & 0.102 & -0.03 & 0.054 & 1.09 & 0.080 & 0.070 & 7.69 & 1.04 & 1.000 & RGB & 0 & 0.24 & 0.440 & 42.4 & 0.7 & 6 \\
HIP6537 & 4744.5 & 50.0 & 2.430 & 0.080 & -0.23 & 0.053 & 1.07 & 0.106 & 0.098 & 6.40 & 1.75 & 0.009 & HB & 0 & 0.16 & 0.373 &   &   & 0 \\
HIP7118 & 4721.8 & 50.0 & 2.501 & 0.098 & -0.18 & 0.056 & 1.46 & 0.219 & 0.186 & 2.64 & 1.82 & 0.272 & HB & 1 & 0.91 & 1.127 & 25.5 & 0.0 & 1  \\
HIP8541 & 4671.7 & 61.2 & 2.945 & 0.130 & -0.19 & 0.056 & 0.96 & 0.059 & 0.041 & 10.97 & 1.31 & 1.000 & RGB & 0 & -0.05 & 0.158 & 50.0 & 0.0 & 6 \\
HIP9313 & 4864.6 & 50.0 & 2.579 & 0.092 & -0.04 & 0.057 & 1.77 & 0.135 & 0.405 & 1.58 & 1.75 & 0.008 & HB & 0 & 0.50 & 0.699 &   &   & 0  \\
HIP9406 & 4943.7 & 50.0 & 3.428 & 0.071 & -0.04 & 0.054 & 1.16 & 0.065 & 0.064 & 5.83 & 0.72 & 1.000 & RGB & 0 & 0.20 & 0.363 & 43.9 & 4.7 & 11 \\
  
\hline\hline                 

\end{tabular}

\end{sidewaystable*}

\begin{figure}
\centering
 \includegraphics[width=0.45\textwidth]{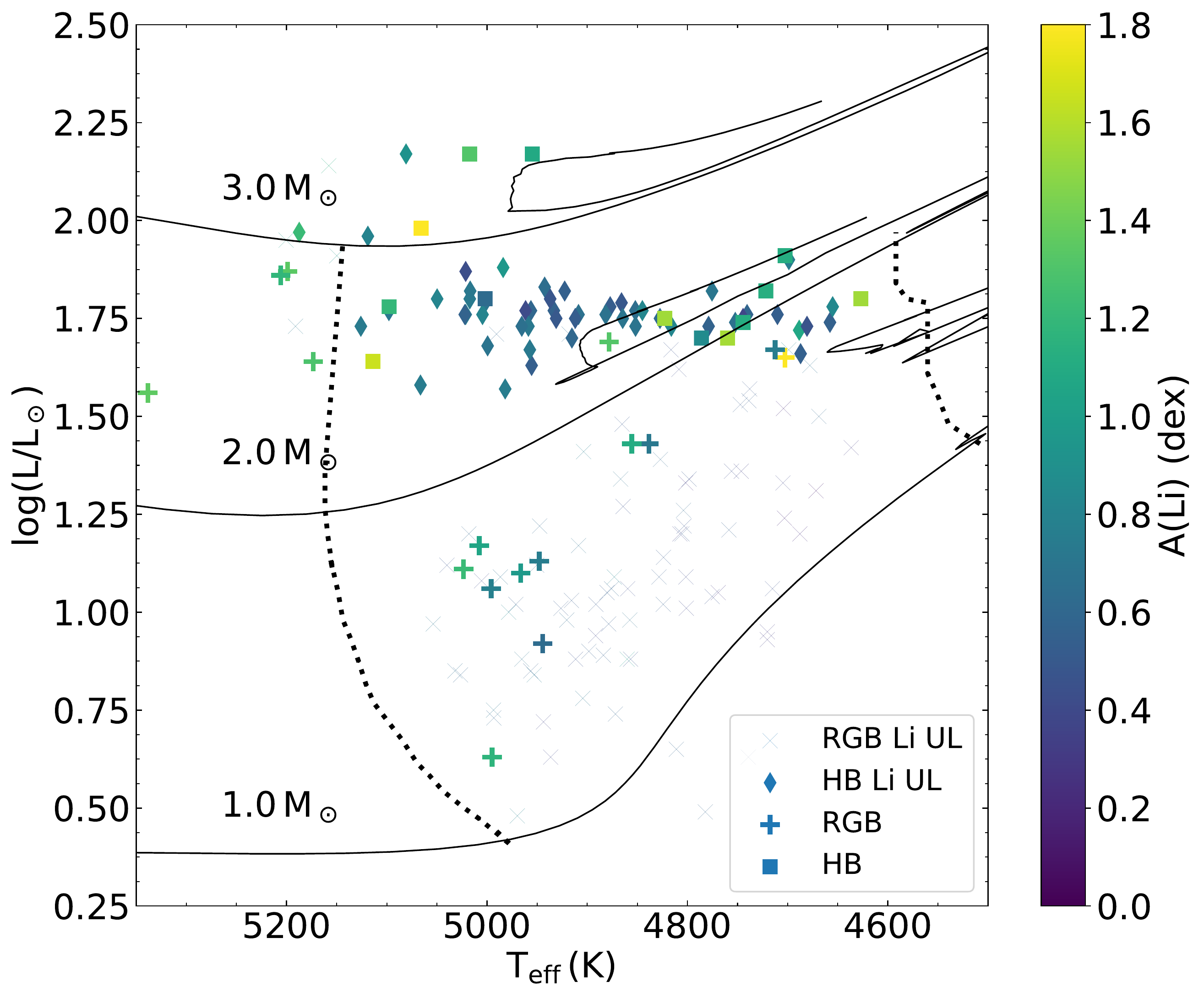}
 \includegraphics[width=0.45\textwidth]{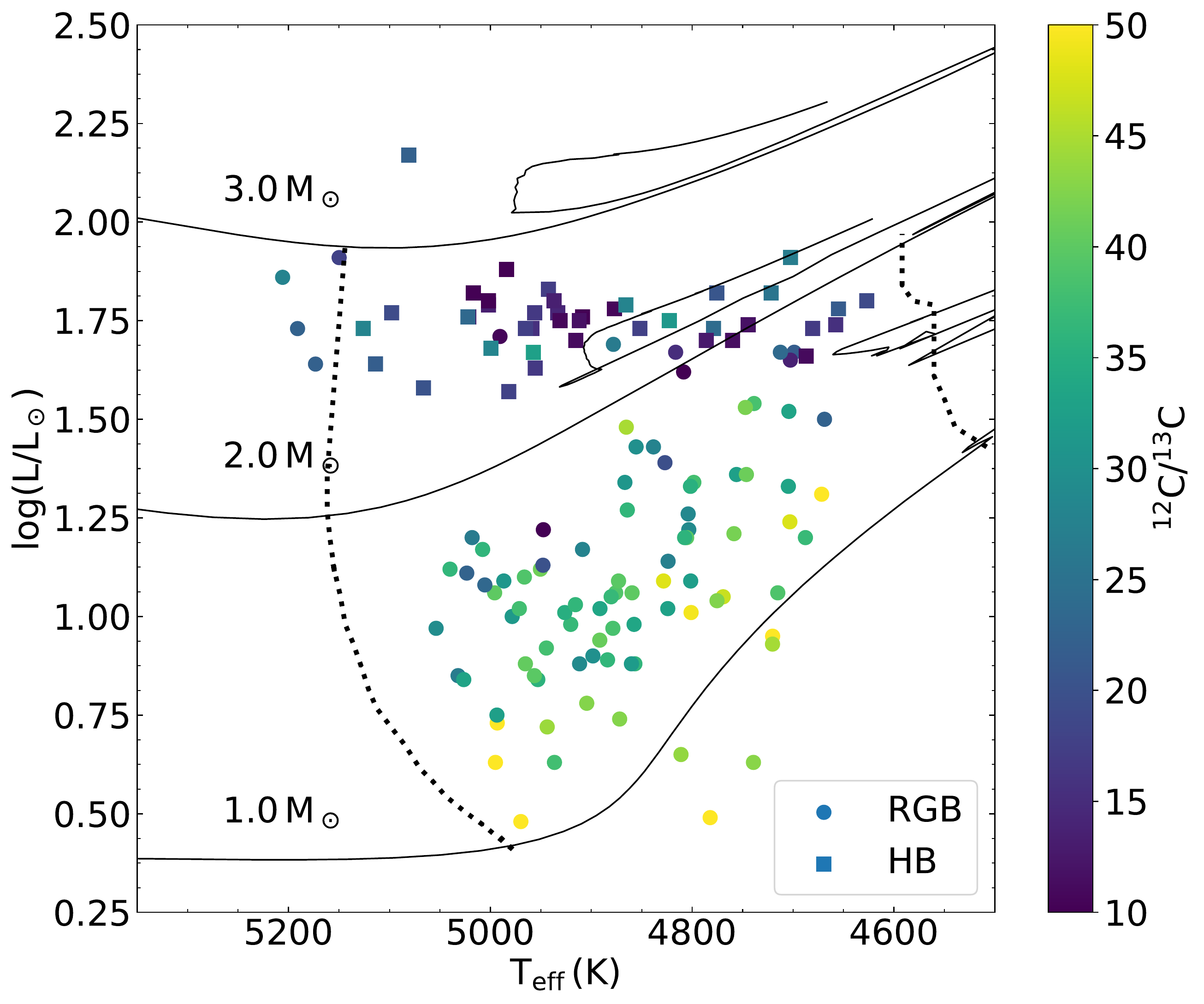}
      \caption{HR-diagram of sample stars. In the top (bottom) panel, they are color-coded by their $A$(Li) (\cir). Different symbols represent stars in different evolutionary stages. Top panel also includes different symbols for Li upper limits and measurements. MESA evolutionary tracks for solar-metallicity with different masses are included. The approximate position of the base of the RGB and luminosity bump is depicted with dotted lines. Note that the Li abundance of HIP3137 significantly exceeds the top of the color bar range.}
    \label{fig:HRdiag}
 \end{figure}

The location of all 166 stars in the HR diagram is shown in Figure \ref{fig:HRdiag}, color-coded by their Li abundances and carbon isotope ratios.
This also includes a classification of their evolutionary status, either in the RGB or the HB. Most of our sample stars do not have asteroseismically-determined evolutionary states, so we base our classification on the probabilities produced by SPECIES (as explained in section \ref{sec:atmospheric_pars}), based on isochrones and the HR diagram. For simplicity, we assumed that all stars with probability $\mathrm{P_{HB}}>0.5$ to be HB. From the 63 stars catalogued as HB, 58 have $\mathrm{P_{HB}}>0.8$, and 51 $\mathrm{P_{HB}}>0.9$. Notice that a subsample of stars with asteroseismic measurements was used in \citep{Soto2021} to validate their methodology. 

Since the large majority of giants in the lower RGB only have upper limits, it is not possible to see the expected drop in A(Li) due to the first dredge-up. The higher A(Li) values tend to be concentrated at the luminosity bump region, where most of our HB stars are located. At the same time, most of the HB stars present only upper limits. Also, although it might seem that stars with higher effective temperatures have higher Li abundances, most of the giants located at high luminosities and higher temperatures have only Li upper limits. This is because it is more challenging to measure reliable Li abundances at higher \teff.

In the case of the carbon isotope ratio, we see that \cir decreases with increasing luminosity.  Given that most of our RGB stars are below the RGB bump, extra mixing has not kicked in yet, and thus this decrease corresponds to the effect on \cir of the first dredge-up.  In our EXPRESS sample, therefore, the evidence for extra-mixing past the RGB bump is provided by the lower average \cir of the HB stars in comparison to the average \cir of the most luminous RGB stars (which are at about the same luminosity as that of the HB).  The low levels of \cir of our HB sample is consistent with measurements in higher luminosity RGB stars and clump stars in clusters \citep[][and references therein]{Maas2019, Lagarde2019}.  This expected decrease in \cir would be an independent way to confirm the evolutionary stage of the giants, as, due to mixing, the \cir is expected to be lower in HB stars. We confirm this throughout this work (See Section \ref{sec:c12c13} and Section \ref{sec:mixing}).

\subsection{Lithium and stellar masses}
 
 \begin{figure}
\centering
 \includegraphics[width=0.45\textwidth]{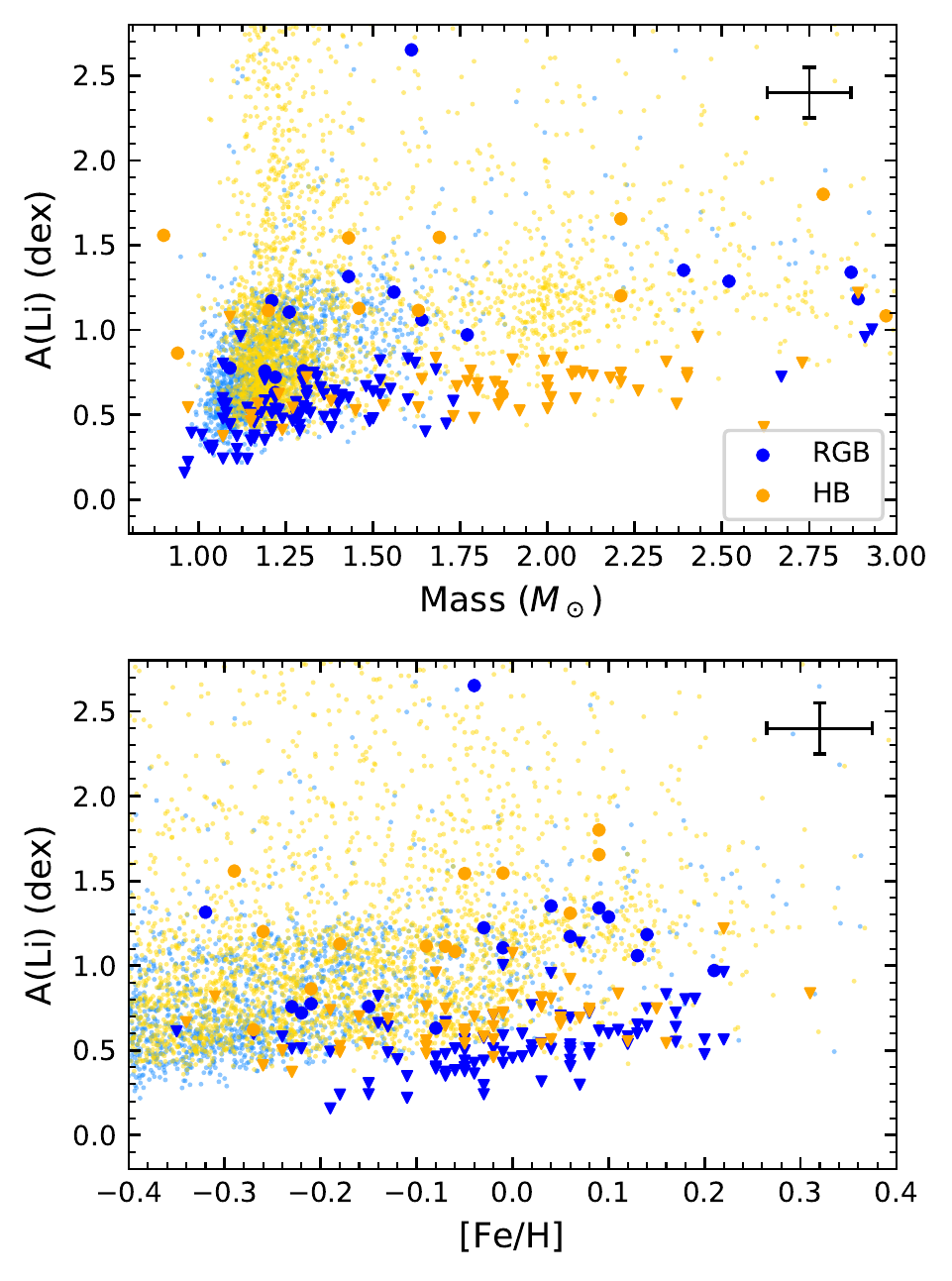}
      \caption{Li abundance and its dependence on mass and metallicity. Stars are color-coded by their evolutionary stage, and GALAH data is included with smaller points to compare our sample. Points are Li measurements and triangles, upper limits.}
    \label{fig:Livsall_limited}
 \end{figure}
 
While the Li abundance of giants depends on several different stellar properties, such as effective temperature, $\log g$, metallicity, their rotation rate, and mass, among others, we concentrate on mass and metallicity, as those are likely the ones that drive most of the phenomenology. We see how Li changes with mass and metallicity in Figure \ref{fig:Livsall_limited}.

Blue points in this figure are RGB stars and orange points HB giants, as identified using isochrone fitting. The downward triangles are Li upper limits.
For comparison, we include as smaller points with the same color-coding the Li abundance measured in the GALAH DR3 catalog \citep{Buder2021}, with masses and evolutionary stages as calculated using the BSTEP code \citep{Sharma2018}.
The GALAH giants are selected by using the criteria presented in \citet{Martell2021}, consistent of cuts because of quality flags and in E(B-V). We consider all stars with $\log g<3.2$, instead of considering a range similar to that of our sample stars ($\log g<3.6$) to minimize the contamination by subgiant stars that have a naturally higher Li abundance.  Also, the GALAH selection considers only measurements and no upper limits. 

Although the Li distributions overlap, HB stars seem to display higher Li abundances than RGB giants. However, to interpret this pattern we require a better understanding of mixing in stars of different masses and metallicities. Notice that while our Li measurements overlap with the GALAH distribution, our Li upper limits are located below the bulk of the GALAH giant distribution with measured abundances, which could be expected, as the upper limits depend on the quality of the spectra.

One of the most important parameters that affects the evolution of stars is their mass. With only small variations in the mass, stars could have vastly different histories: giants with masses of $\sim1.3\ \mathrm{M_{\odot}}$ may have evolved from the Li dip and could have depleted their Li abundances during the main sequence. In contrast, stars with higher masses are expected to preserve their original Li abundances, and stars with lower masses may also be Li-depleted by extra-mixing processes acting on the main sequence. Then, the mass is also relevant to define the expected Li abundance pattern of normal giants in the population, which allows to precisely identify Li-enriched objects. In \citet{AG2016}, after modeling the process of sub-stellar companion engulfment by giants of different parameters, the authors conclude that the limit that is commonly used to define a Li-rich giant ($A\mathrm{(Li)}>1.5$ dex), i.e., a giant that suffered non-standard evolution, is not always applicable. \citet{Kumar2020} also conclude that the limit to identify a Li-rich giant is not correct, but now considering that most of them are core-He burning. They suggest that all core-He burning giants are going though a process of enrichment, and thus, they are all enriched. This statement is strong and has been debated. Using open cluster data from the Gaia-ESO Survey, \citet{Magrini2021} find that, although a Li-enrichment process may be acting in some of the clump giants, this is not needed for all of them. This is not the first time such a statement, about a possible ubiquitous enrichment of red clump stars, has been made, with subsequent evidence disfavoring the scenario \citep[See][]{Carlberg2016}. By considering not only the evolution of stars of different masses, but also the differences between the red clump and red giant populations found in the GALAH sample, \citet{Chaname2021} also conclude that a Li-enrichment process is not mandatory to explain the abundances of all core-He burning giants.

Thus, mass matters. And, to properly characterize the Li pattern of field stars and select giants that are enriched, the safest way is to compare between stars of similar masses and metallicities, and identify giants with higher abundances. We show the Li pattern of giants with mass in the upper panel of Figure \ref{fig:Livsall_limited}. There is a slight correlation between Li and mass in the meassurements. There are HB giants all across the mass range, but we see an absence of RGB stars between masses of $1.7$ to $2.5\ \mathrm{M_\odot}$. These mass distributions come from our sample selection, and we do not expect our stars to be representative of the entire population of RGB and HB stars. In general, the peak of the mass distribution for both HB and RGB stars in the neighborhood is close to $1.2\ \mathrm{M_\odot}$, with a more rapid drop off in the number for RGB stars. In general, there are more massive HB stars (masses larger than $1.8\ \mathrm{M_\odot}$) than RGB giants, and we see that too in the comparison GALAH distribution. Given that stars of different masses can have different histories of Li depletion, this difference in the distribution should be considered when analyzing Li abundance. The subset of core-He burning giants that are more massive do not experience a process of Li depletion in the main sequence, and thus, in general, we can expect a higher Li abundance in this population. With the mass range of the RGBs, it is possible that they come from below or from the Li dip. 

In the second panel, we see no correlation of $A$(Li) with metallicity, and no difference in the distribution between first ascending RGB and HB giants. The mean metallicity of the sample is solar \citep[See ][]{Soto2021}, with no selection in favor of metal-rich stars. We note that there seems to be more RGBs with measurements (instead of upper limits) above solar metallicity. HBs with higher Li also seem to concentrate around solar metallicity.

Selecting exactly what is a Li-enriched giant might be complicated. The distribution of abundances for the normal giants must be properly characterized so that outliers can be detected. However, a simple estimate based on the giants that are above the distribution would indicate that HBs are much more likely to be enriched than RGBs. This qualitatively agrees with \citet{Martell2021}. We do not calculate the probabilities of enrichment for our giants because our sample considers a previous selection of targets based on magnitude that would bias the calculation, and second, because a limiting $A$(Li) that does not consider mass could be misleading \citep{AG2016}. Additionally, there are only 5 stars in our sample with abundances $A$(Li)$>1.5$ dex, the typical limit used to identified Li-rich giants, which makes calculations of rates of enrichment depending on different parameters impossible.

\begin{figure*}
\centering
 \includegraphics[width=0.6\textwidth]{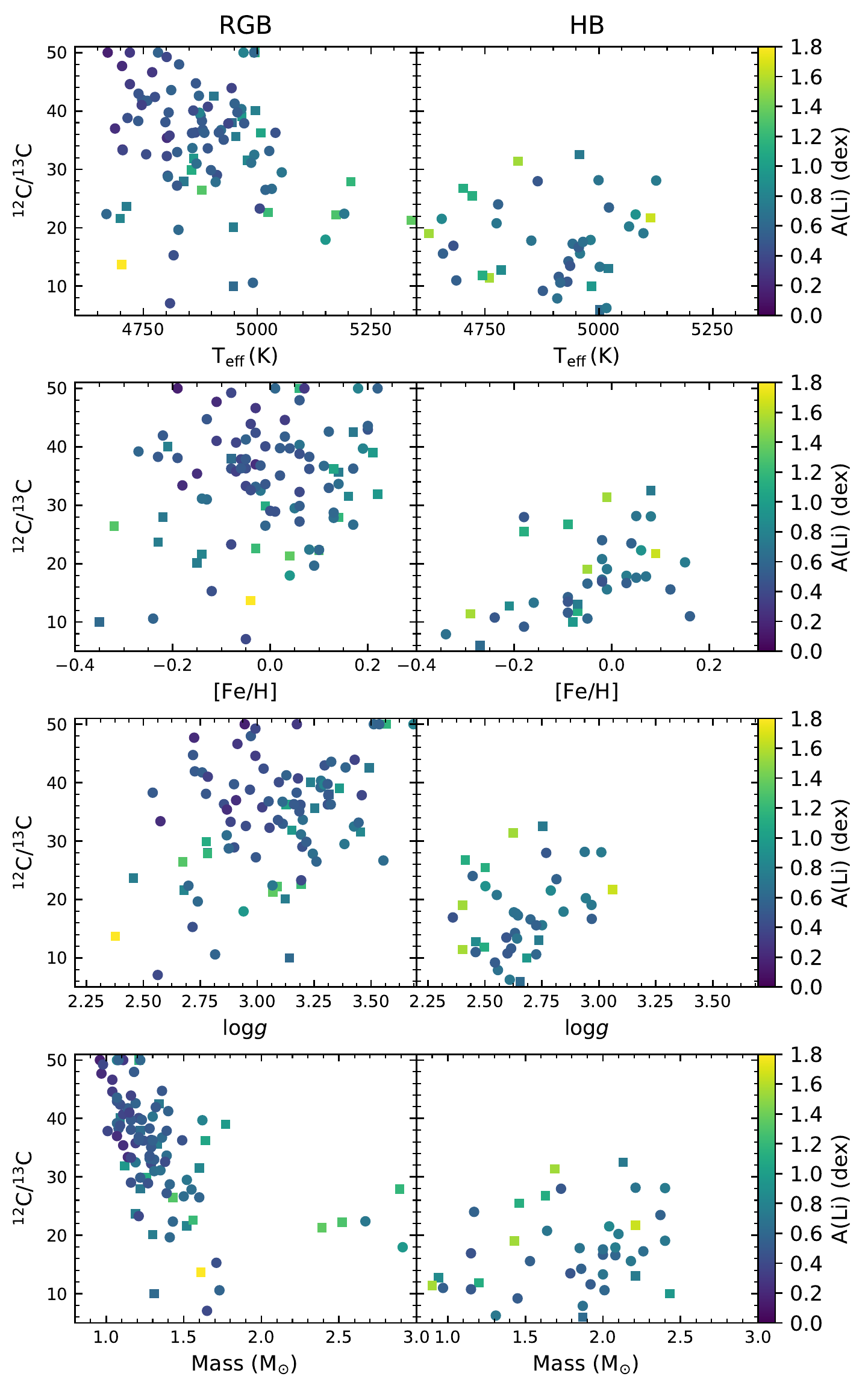}
      \caption{Carbon isotope ratios \cir of sample stars in different evolutionary stages (RGB left panels, HB right panels) against different stellar parameters. Additionally, stars are color-coded by their $A$(Li), with different symbols indicating Li upper limits (circles) and measurements (squares). The $A$(Li) range of the color bar is selected to make variations of most of the sample stars clearer, but $A$(Li) of HIP3137 exceeds the top value of this range.}
    \label{fig:c12c13_vs_all}
 \end{figure*}
 
We can identify some giants in this figure that can be classified as Li-rich when compared not only to the rest of our sample, but also to the GALAH comparison sample. The two horizontal branch stars with lowest mass (HIP63583 and HIP60396), have a higher Li abundance than the rest of the sample. Also, there are no HB stars in the GALAH comparison sample at these low masses. These two stars are classified with a high probability as core-He burning and their \cir is consistent with that evolutionary phase with \cir$=11.4$ for HIP63583 and \cir$=12.8$ for HIP60396 (see Section \ref{sec:c12c13}). Even if this classification is incorrect, HIP63583, with $A$(Li)$\mathrm{_{NLTE}}=1.56$ dex still has a higher Li than the bulk of the population at the same mass.

In the mass range $1.1-1.5\ \mathrm{M_\odot}$, the giants with Li detections in our sample have a Li abundance within the limits of the bulk of the GALAH giants. Thus, they appear to have a normal Li abundance, with the possible exception of the clump star HIP5364 and the RGB star HIP70261. 
For masses higher than that, we first identify the giant with the largest Li abundance in our sample, HIP3137. With $A$(Li)$\mathrm{_{NLTE}}=2.65\pm0.16$ dex, it is located well above the bulk of the population. Notice also that the most Li-enriched RGB star in our sample is also the star with the lower $\log g$.
Stars like HIP76532, with a mass of $1.69\pm0.21 \mathrm{M_\odot}$ and $A$(Li)$\mathrm{_{NLTE}}=1.55\pm0.14$ dex and HIP110529, with a mass of $2.21\pm0.11 \mathrm{M_\odot}$ and $A$(Li)$\mathrm{_{NLTE}}=1.66\pm0.13$ dex, also have Li abundances higher when compared to other stars of similar masses. Additionally, with a mass of $1.56\pm0.09 \mathrm{M_\odot}$, the RGB star HIP77059 has $A$(Li)$\mathrm{_{NLTE}}=1.22\pm0.13$ dex, at the limit of the normal distribution.

At higher masses  (M$\gtrsim 2.2\ \mathrm{M_\odot}$), there are less comparison stars, but mixing is fundamentally different, given that the H-burning shell never reaches the discontinuity left by the convective envelope before it evolves off the RGB. Thus, it is possible that Li is naturally higher both in the RGB  after the first dredge up, because of the shallower depth of the convective envelope, and upper RGB and HB stars. The observed scatter could be just limited to different initial Li abundances (and mixing in the main sequence) and different depletion rates at different metallicities. 

\subsection{Carbon isotope ratio and mixing} \label{sec:c12c13}

The carbon isotopic ratio is also an important tracer of extra-mixing processes, and can depend on evolutionary stage, mass, and metallicity. We check if there is any clear dependency of \cir with stellar parameters and with Li abundance in Figure \ref{fig:c12c13_vs_all}.  
This has been separated into RGB (left) and HB (right), and color-coded by their Li abundances.

The first thing we notice is that, as expected, the HB tend to have a much lower \cir than RGB stars. This indicates that, overall, the determination of evolutionary stages by \citet{Soto2021} is correct. HB tend to have lower \cir, given that extra-mixing is expected to decrease this ratio as the star ascends the RGB. This is consistent with the behavior of thermohaline mixing in the RGB according to theoretical models \citep{Lagarde2019}. In \citet{Shetrone2019}, the authors find that at the metallicity of our stars, there is no evidence of extra-mixing based on [C/N] abundance ratios. However, \citet{Fraser2022} find some unrealistic unmixing in the APOGEE measurements an effect that canbe seen in the higher metallicity bins. Correcting by that effect, they do find small amounts of extra-mixing at high metallicities which is consistent with what we find in this work.

In the RGB phase, there seems to be a slight decreasing correlation with mass (bottom panel Figure \ref{fig:c12c13_vs_all}). 
Lower mass stars tend to have higher \cir, while up to 2.0 $\mathrm{M_\odot}$ this decreases steadily. The stars at higher mass, of $\sim 2.5\ \mathrm{M_\odot}$ do have a higher \cir, which is consistent with standard stellar evolution models and observations in clusters \citep{Gilroy1989}. 
Most of the RGB stars in our sample seem to be at the base of the RGB, thus, it is possible that some have not even completed their first dredge up, consistent with their higher $\log g$ values. Although these stars have very low Li abundances, which could suggest that they have already depleted their Li, it is also possible that they have very low Li abundances previous to their RGB phase. No thermohaline mixing is expected yet in these stars.

The observed trend in mass for RGBs is somehow consistent with expectations from standard theory, with the scatter indicating both the presence of main sequence mixing and a first dredge up still in progress. Standard theory indicates that the location of the $\mathrm{^{13}C}$ peak found in the interior of the star changes with mass, being closer to the surface for higher masses. Thus, more $\mathrm{^{13}C}$ can be brought to the surface for in higher mass stars decreasing their \cir value post first dredge up.

\begin{figure*}
\centering
 \includegraphics[width=0.8\textwidth]{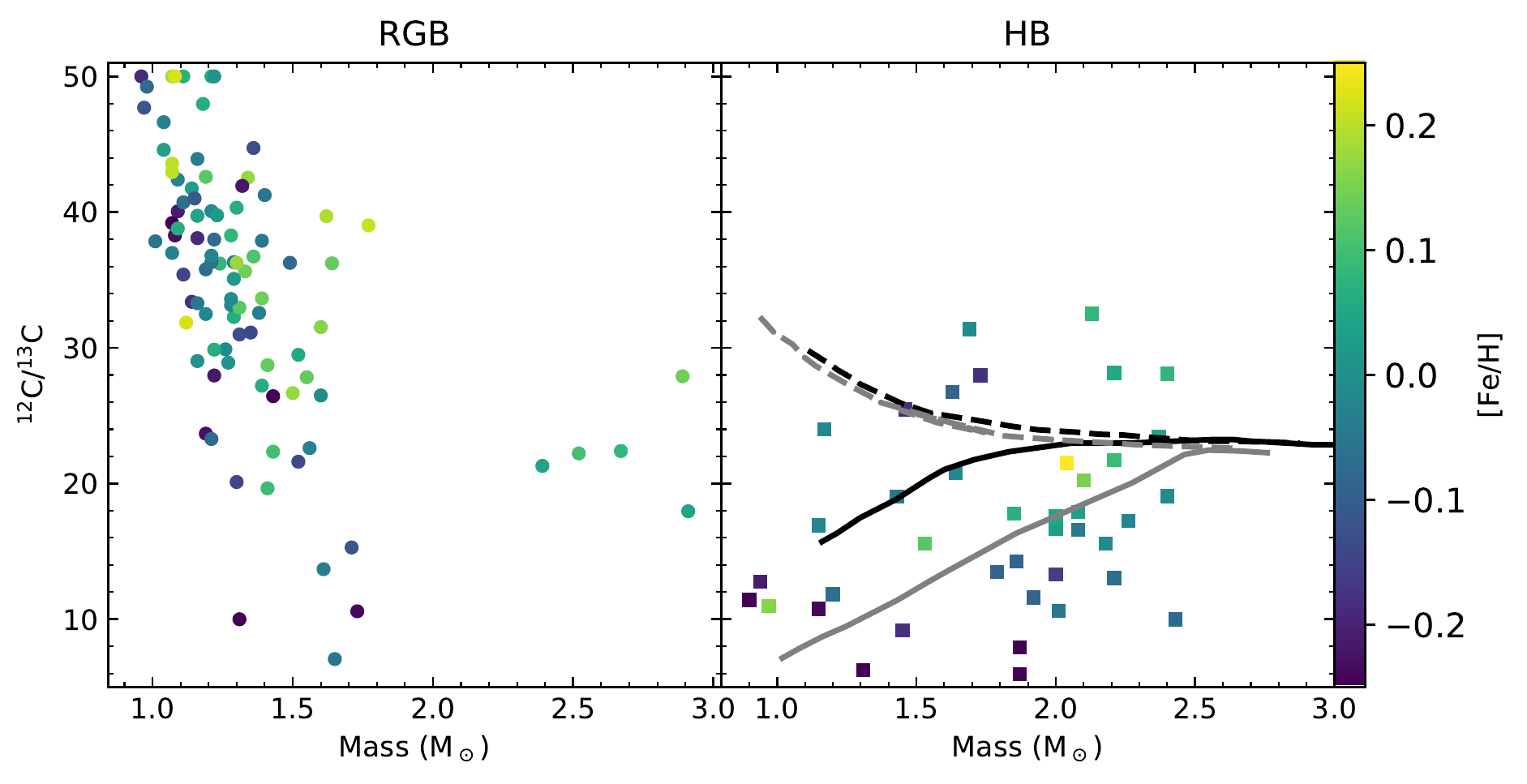}
      \caption{Carbon isotopic ratio as a function of stellar mass, color-coded by the metallicity. Left (right) panel considers RGB (HB) stars. Additionally, the locus of synthetic thin disk populations from \citet{Lagarde2017} are included in the panel for core-He burning giants. Solid lines are with thermohaline mixing and dashed line for giants with no extra-mixing. Black lines inidicate the position of modeled stars of [Fe/H]$\sim0.5$, while gray lines are for stars of [Fe/H]$\sim-0.5$.}
    \label{fig:c12c13vsmass_met}
 \end{figure*}

Another very interesting feature in Figure \ref{fig:c12c13_vs_all} is the \cir of HB stars, that seems to increase slightly for higher mass stars. This behavior is also consistent with thermohaline mixing. The shape of the distribution is very similar to that presented by \citet{Lagarde2017}, although those giants were in a different metallicity range. A comparison with the synthetic population of that work can be found in Figure \ref{fig:c12c13vsmass_met}.

The \cir of the HB giants also has a strong positive correlation with metallicity. Stars at solar and super-solar metallicity have a larger scatter, with \cir from $\sim18$ to $30$. In contrast, stars of lower metallicites tend to cluster around $\lesssim10$. RGB stars do not show such a dependency on metallicity. This is also consistent with a metallicity dependent extra-mixing mechanism acting on the RGB, being more efficient for metal-poor stars and thus further decreasing their \cir.

To analyze the possible effect of transport processes in the RGB we add to the \cir information the Li abundance, represented by different colors in Figure \ref{fig:c12c13_vs_all}, as well as symbols, with squares being measurements and points upper limits. 

Although we could have expected that HB stars had a much lower Li abundance if they went through extra-mixing, the large amount of upper limits does not allow to conclude directly about the $A$(Li) of evolved giants. On the other hand, the presence of a few HB stars that have a higher abundance would be consistent with the idea that some giants can experience Li production in the He-flash \citep{Magrini2021,Chaname2021}, although most likely not all of them as suggested by \citet{Kumar2020}.

\begin{figure}
\centering
 \includegraphics[width=0.45\textwidth]{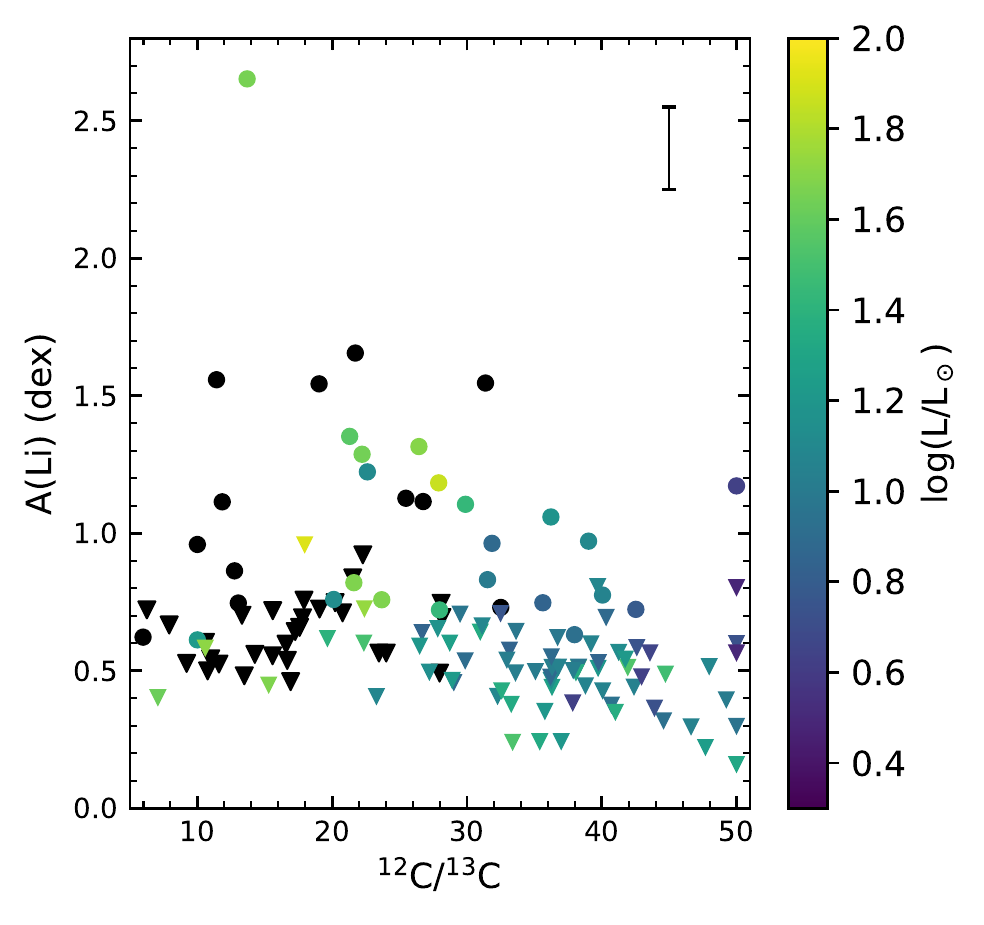}
      \caption{Correlation between Li abundance and \cir for RGB (colored symbols), color-coded by their luminosities, and HB giants (black symbols).}
    \label{fig:c12c13vsli_lum}
 \end{figure}

There is no other evident correlation between parameters including both the abundances of Li and \cir. We check additionally in Figure \ref{fig:c12c13vsli_lum} changes in both $A$(Li) and \cir with luminosity. The carbon isotopic ratio should decrease as the star increases the luminosity, first, because of the first dredge-up, and after the luminosity function bump, due to the effects of extra-mixing. We see a similar trend in our data, with lower \cir at higher luminosities, although our RGB stars have not yet started their Li depletion after the bump, and thus the effect is most likely the effect of the first dredge-up. Notice that in this figure, clump stars are colored in black not to confuse them with first ascending RGBs.

\section{Discussion: Extra-mixing in the RGB} \label{sec:mixing}

Extra-mixing in the RGB depends on several different stellar properties. Studies in clusters have suggested that extra-mixing is efficient even at higher metallicities \citep{Szigeti2018}, but the mass is something that needs to be considered before claiming a certain correlation with parameters. Given the metallicity selection of our sample, this is something that we cant test in this work. Figure \ref{fig:c12c13vsmass_met} shows the \cir as a function of mass, color-coded by the metallicity and separated into different evolutionary stages. There is a correlation with metallicity for the HB stars, with metal-poor stars showing lower ratios at fixed masses.

Theoretical models that include thermohaline mixing \citep{CharbonnelLagarde2010, Lagarde2019}, indicate that the mixing efficiency increases in less massive stars at a given metallicity, or increases in more metal-poor stars at a given stellar mass. This seems to be consistent with the overall trends we find in this work. A direct comparison between population models in \citet{Lagarde2017} and observations (Right panel Figure \ref{fig:c12c13vsmass_met}) shows that measurements are not consistent with theoretical predictions with no extra-mixing (dashed lines). Instead, predictions including thermohaline mixing better explain HB stars with low \cir, although probably additional ingredients must be included to explain the scatter, such as rotational mixing.

Models \citep{Lagarde2012} show that during the first dredge-up, higher mass stars have a larger \cir depletion, with little effect of the metallicity. However, after thermohaline extra-mixing proceeds, this correlation is reversed, and the \cir decreases much more for the lower-mass stars. 

\begin{figure*}
\centering
 \includegraphics[width=0.9\textwidth]{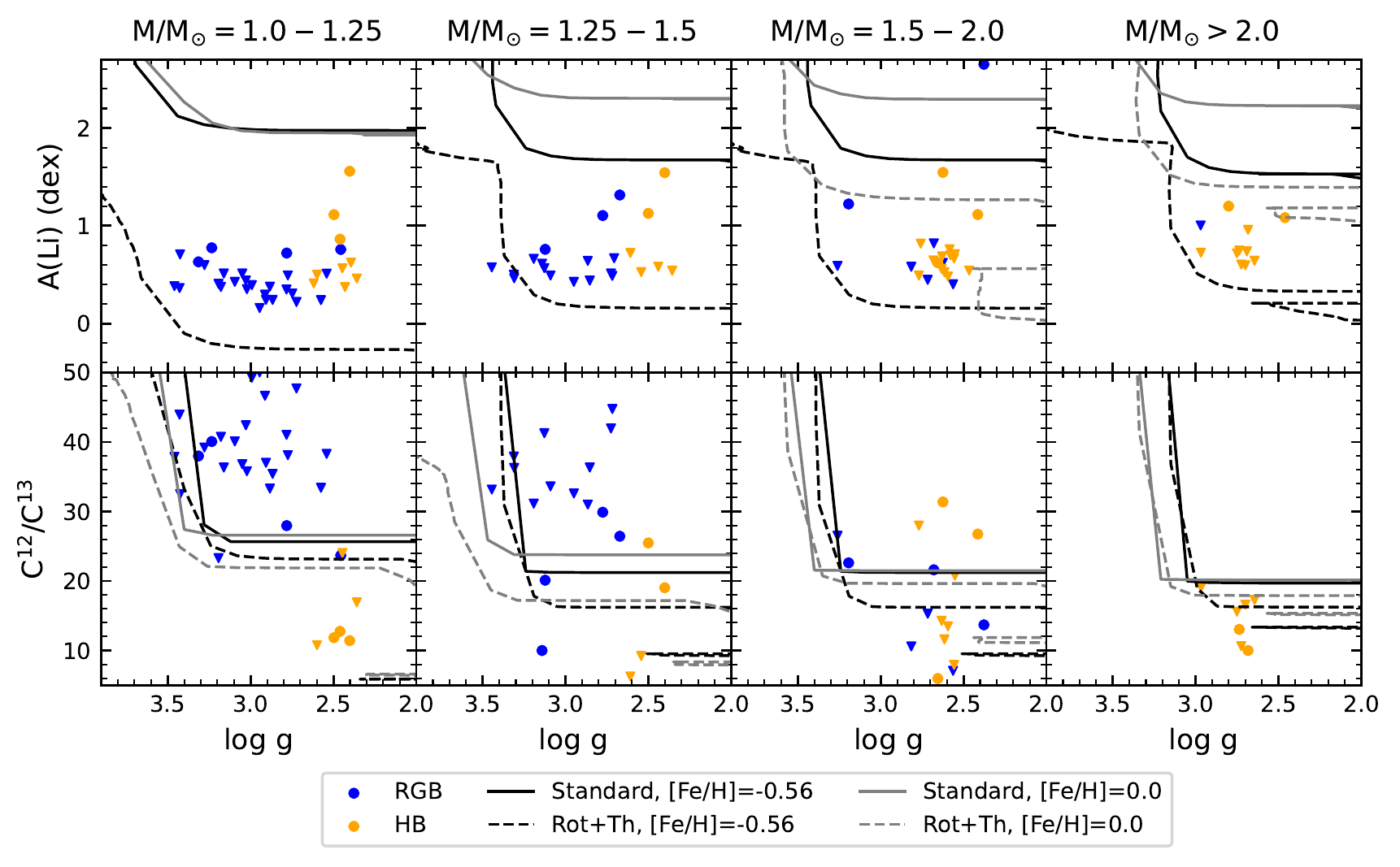}
      \caption{$A$(Li) (upper panels) and \cir (lower panels) of stars with [Fe/H]$<0.0$, binned in mass, as a function of $\log g$. Lines are theoretical models from \citet{Lagarde2012}. Solid lines are standard models, while dashed lines are models that consider both rotational and thermohaline mixing. Masses for the models are $\mathrm{M/M_\odot}=1.0, 1.5, 1.5, 2.0$ from left to right panels. Triangle symbols in the lower panels are Li upper limits.}
    \label{fig:vslogg_metpoor}
 \end{figure*}
 
 \begin{figure*}
\centering
 \includegraphics[width=0.9\textwidth]{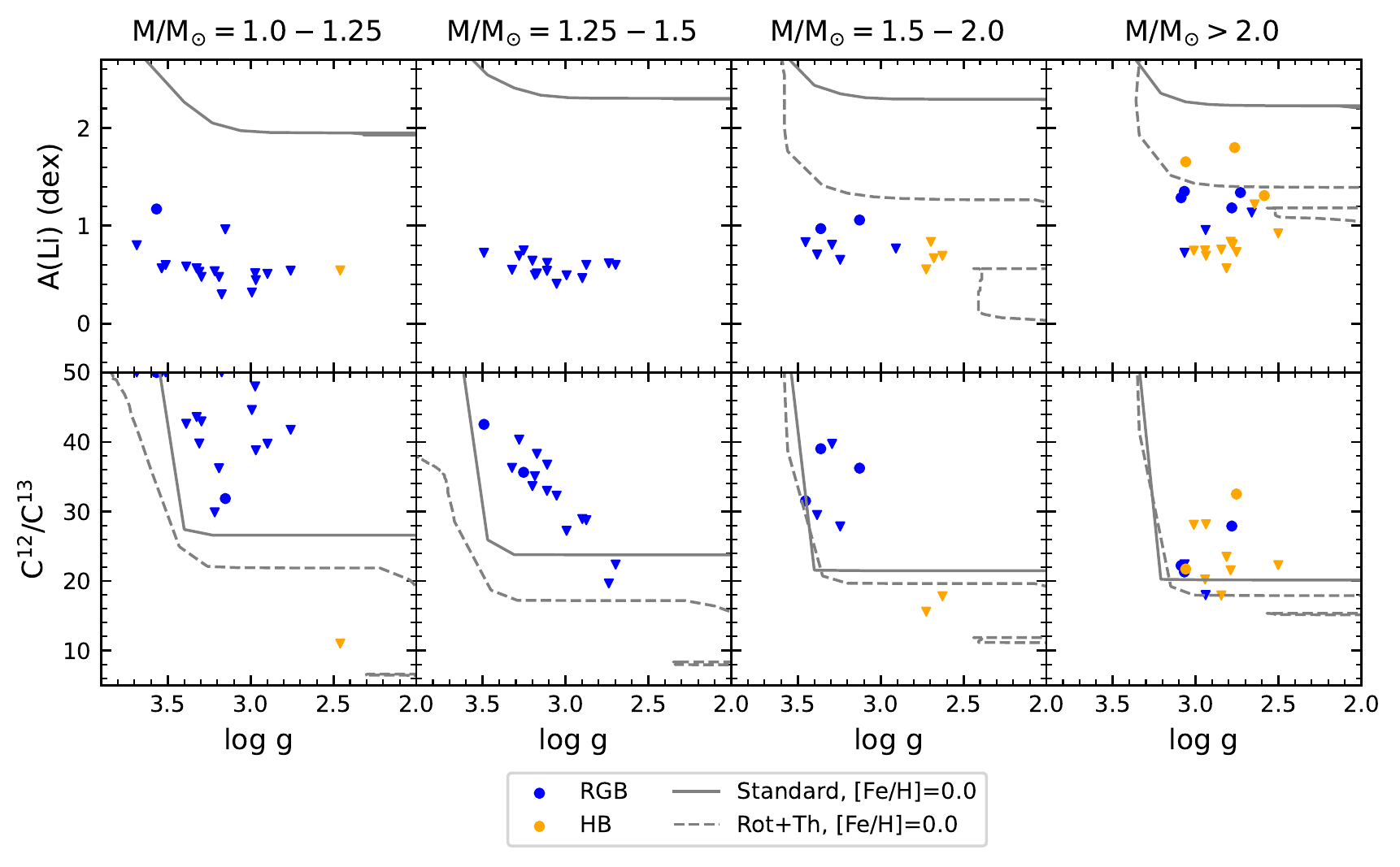}
      \caption{$A$(Li) (upper panels) and \cir (lower panels) of stars with [Fe/H]$\geq 0.0$, binned in mass, as a function of $\log g$. Lines are theoretical models with solar metallicity from \citet{Lagarde2012}. Solid lines are standard models, while dashed lines are models that consider both rotational and thermohaline mixing. Masses for the models are $\mathrm{M/M_\odot}=1.0, 1.25, 1.5, 2.0$ from left to right panels. Triangle symbols in the lower panels are Li upper limits.}
    \label{fig:vslogg_metrich}
 \end{figure*}

To analyze these trends in the data, we attempt to emulate a science case similar to that of star clusters by separating the sample stars into two metallicity ranges, and in several different mass bins. That allows to follow the evolution of similar stars (although the range in metallicity is relatively large) and enables a cleaner comparison to models. Figures \ref{fig:vslogg_metpoor} and \ref{fig:vslogg_metrich} show the Li (top panels) and \cir (bottom panel) of metal poor and metal rich stars respectively, binned in 4 mass bins. In the last bin (rightmost panel) some stars will not experience mixing between the RGB and HB phases (i.e. masses higher than $2.2\ \mathrm{M_\odot}$).

The carbon isotopic ratio in most cases seems to be lower in HB stars than in RGB stars of similar stellar properties, which confirms an important depletion factor during the evolution of the star, when mixing is acting. Also, the ratio in the core-He burning phase is mostly consistent with models from \citet{Lagarde2012}, which consider the evolution of the star from the pre main-sequence up to the asymptotic giant branch accounting for thermohaline and rotationally-induced mixing. We can see the HB Li abundances or \cir of models as horizontal lines at low $\log g$. Li is harder to understand than \cir, with the large amount of stars only showing upper limits. The upper limits seem to be consistent with models and there is not a significant difference between the RGB and HB upper limits. The measurements are usually located between the standard models and those including mixing, a region that can be covered by considering different rotation rates in the models. There are two HB stars at low mass and low metallicity that have a higher Li abundance than other similar stars and this could be an indication of Li enrichment. These two stars, however, seem to be unusually under-massive for HB stars and other possibilities, such as wrong evolutionary phase determination, mass determination, or interaction with a companion should be considered. These do not seem to have either binary companions or orbiting planets.

The other giant that clearly stands out due to its high Li abundance, HIP3137 has a low \cir, which could be an indication that it has gone through efficient extra-mixing, an hypothesis somehow consistent with its $\log g$. This is an RGB star close or past the luminosity function bump. If its evolutionary stage is confirmed, it could have just been recently enriched. Given that planet engulfment is not expected to decrease the carbon isotopic ratio of the star, this seems an unlikely explanation for the abundance of this particular giant. This Li-rich giant is an excellent example of why \cir should definitely be considered when discussing the source of Li-enrichment and anomalous surface abundances in evolved stars \citep[e.g.,][]{Sun2022}.

The amount of HB stars in the higher metallicity bins (Figure \ref{fig:vslogg_metrich}) is much lower. Thus, it is difficult to say, without a doubt, that these higher metallicity stars go through extra-mixing, or attempting to constraint the efficiency of this mixing.
Instead, for the sample between -0.25<[Fe/H]<0 (Figure \ref{fig:vslogg_metpoor}), we can more clearly see that mixing is acting when we compare the abundances of RGB and HB giants. In these figures, we have considered only two broad metallicity ranges, and thus cannot evaluate if the mixing is metallicity dependent. However, Figure \ref{fig:c12c13vsmass_met} seems to suggest that lower metallicity HB stars have lower \cir values, but solar-metallicity stars still show the effect of extra-mixing, with lower abundances in the HB than in the RGB before the bump.

Other studies have analyzed the presence of extra-mixing and its dependence on stellar parameters. In \citet{Gratton2000}, they study stars with lower metallicities, from -2 to -1 dex, looking for signatures of transport processes on Li, C, N, and \cir, but their sample size does not allow to understand how mass and metallicity affect the extent of extra-mixing.
Additionally, \citet{Shetrone2019} study the effect of extra-mixing in alpha-rich field giants traced by [C/N], using APOGEE and the large number statistic that this survey provides. Their alpha-rich population (limiting the mass of the sample to lower values $\sim1.0\ \mathrm{M_\odot}$) has metallicities from -1.7 to 0.1 dex. This wide range allows them to conclude that the post-luminosity bump mixing is strongly metallicity dependent and grows as metallicity decreases, starting from [Fe/H]$\sim-0.5$ dex. Notice that their predictions of no mixing at higher metallicity might be due to problems with the C and N measurements \citep{Fraser2022}.

Our sample, centered at solar metallicities, shows the overall effect of extra-mixing between the lower RGB, before the luminosity function bump, and the core He-burning phase. By using the Li and \cir, we find that the prescription with extra-mixing fits better the data than standard models.   
The presence of mixing at higher metallicities acting on \cir is consistent with results from open clusters \citep{Gilroy1989, Szigeti2018}.

Various light elements might trace transport processes differently, depending on the depth of the mixing. 
An homogeneous sample with measurements of different elements, following the entire RGB evolution would help to understand better the depth of any mixing process acting after the RGB luminosity bump and calibrate its efficiency to fit the evolution of all chemical species at the same time \citep[e.g.][]{Maas2019, TayarJoyce2022}.

\section{Binarity and presence of planets} \label{sec:binary}

Some models of Li-enrichment require the giant to have a binary companion. \citet{Casey2019} suggest that tidal spin-up from a binary companion could drive extra-mixing inside the star, increasing the Li abundance in the surface. Given that our sample is part of a planet search program, these stars have multiple radial velocity epochs that can point to possible binaries.

We analyze if the Li-enrichment is associated with the presence of planets or binary companions, and if we can find any trend with carbon isotopic ratios. Spectroscopic binaries in our EXPRESS sample are reported in \citet{Bluhm2016}, while planet detections have been published in several different works (e.g. \citealt{Jones2013}, \citealt{Jones2016}, \citealt{Jones2021}).

\begin{figure*}
\centering
 \includegraphics[width=0.65\textwidth]{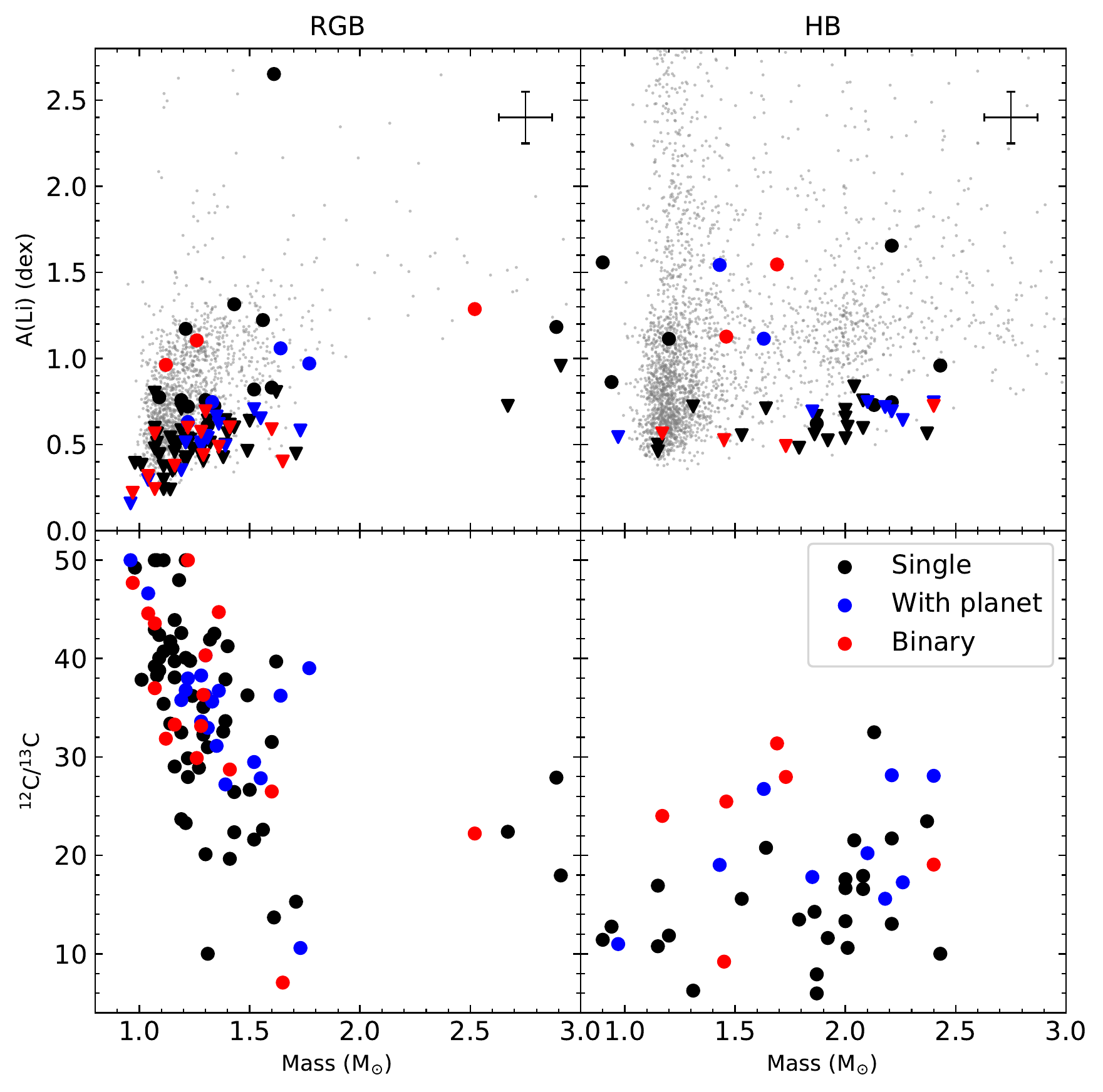}
      \caption{Li (top panels) and carbon isotopic ratio (lower panels) with mass for giants in the RGB (left) and HB (right). Stars are color-coded according to their status as single, with planets, or binary.}
    \label{fig:binaryplan}
 \end{figure*}

We find no correlation between high Li abundances and the presence of a binary companion or planet in general.
Most of the stars with either planets or binaries have $A$(Li) upper limits, or measurements consistent with the GALAH sample (see top panels of Figure \ref{fig:binaryplan}, separated into RGB and HB). Notice that if the star already engulfed a planet, we do not necessarily need to find additional planets orbiting the host star, and thus, the lack of planetary companions does not give additional information about the mechanism of Li enrichment. Although some binary stars such as HIP76532 and stars hosting planes, such as HIP5364, can have high Li abundances, a general trend does not seem to exist, and it might be necessary to analyze each of these cases individually.

The RGB star with mass $\sim2.5\ \mathrm{M_\odot}$ and $A$(Li)$\mathrm{_{NLTE}}=1.29\pm0.15$ dex (HIP68099) is also a binary. Although its Li abundance is not particularly high, this star and others at higher masses can be interesting, given that at these masses no luminosity function bump is expected. The lack of a larger amount of stars to compare also makes it harder to identify if these giants are unusual.

Stars with low masses and relatively high \cir seem to have companions. Although there are only 4 stars in this category, it is possible that the presence of binaries could also modify the carbon isotopic ratio, either through enhanced mixing, transfer, or any other mechanism, but since there is no clear correlation, it is not possible to derive a clear conclusion from this sample of stars. Certainly a larger sample of known RGB binary systems (e.g. \citealt{Massarotti2008}; Uzundag et al. 2022) should be analyzed to study the effect of binarity on the \cir values. Moreover, no parameters for the secondaries can be extracted with the radial velocity information alone. Most of the stars have degenerate orbital solutions, not allowing to obtain masses or separations for the targets. Analyzing if more massive or closer companions can induce tidal mixing is something more to consider for future studies.

\section{Summary} \label{sec:summary}

The Li abundance of giants can give clues about the nature of extra-mixing in the RGB, a process that changes the surface abundance of light elements and acts efficiently after the luminosity function bump. Additionally, the high Li in some small percentage of giants also reveals unconstrained mass-dependent processes that can affect the abundances of stars. For this, there has been extensive works to understand the Li pattern and nature of the enriched objects, such as using asteroseismology to accurately determine the age of giants, studying different populations such as clusters, or better identifying outliers and constrain the physical mechanism of mixing, among others. 

In this work, we add the analysis of the carbon isotope ratio to the discussion of Li abundances. The \cir is actually one of the best indicators of transport processes in giant stars, but there are only a handful of works with determinations of the ratio in field giants \citep[e.g.][]{Gratton2000,Keller2001,Carlberg2012,Tautvaisiene2013, Takeda2019}. Given that Li is very easily destroyed and depleted, and several factors can alter its abundance, it can be complicated to analyze it properly. \cir is complementary to the Li abundance. With different elements burning at different temperatures, Li and \cir probe different depths inside giant stars allowing to characterize the mixing.

We measure the \cir using the $\mathrm{^{12}C/N}$ and $\mathrm{^{13}C/N}$ bands at $\sim 8000$ \AA. These can be significantly affected by tellurics. Our sample has the additional advantage that it is part of a planet-search program in giants, EXPRESS, which allows to select the epochs where the telluric lines are not affecting the \cir measurement. The presence of binaries and planets has also been associated with the complicated Li abundance pattern of giants. Given the nature of our targets, extensively followed to search for planets, we can also study possible trends between the Li, \cir, and companions. There does not appear to be any obvious correlation between the high Li abundance or \cir and binarity.

We conclude that the \cir\, provides additional clues about the physical mechanism behind extra-mixing. 
For the horizontal branch stars, there is a strong correlation between \cir and metallicity. For these same stars, there is also a correlation with mass. {Both of these trends are more consistent} with population models of stars that include thermohaline mixing than with canonical models that do not. In the RGB, there is an anticorrelation between \cir\, and mass. Given that our RGB stars are located before the bump, with no thermohaline mixing affecting them, we expect this correlation to be relatively narrow and due to the effect of first dredge up.  Thus, overall, the \cir in our sample is consistent with the presence of thermohaline mixing acting in the RGB.

Some stars appear to be unusually enriched in Li. Two of these are HB stars with very low masses and low \cir, consistent with their evolutionary phase. The most Li-enriched giant is classified as RGB, with a mass close to $\mathrm{1.6}\,\mathrm{M_\odot}$, solar-metallicity, and relatively low \cir. It is very close to the luminosity function bump, and it does not show either the presence of a binary companion or planet. There is no clear explanation behind the Li abundance of this giant. Its carbon isotope ratio may give clues about its enrichment process. If it is located after the luminosity function bump, it might be affected by enhanced extra-mixing. Given its low \cir, it is also possible that it is wrongly classified as a RGB. Any external mechanism should not change significantly the \cir, so this does not seem to be the explanation for the high Li of this particular giant.

The \cir\, has proven to be a very useful tool when studying extra-mixing or the enhanced Li of RGB stars. More measurements in field stars, hopefully with known masses, are needed to provide better comparisons with models and to better understand transport processes inside giants.

\begin{acknowledgements}
We would like to thank the referee for their suggestions, that improved the quality of this work. C.A.G. acknowledges financial support from ESO-Chile Joint Committee and the Millenium Nucleus ERIS NCN2021\_017.  C.A.G. woud like to thank J. Tayar for useful comments.  J.C. acknowledges support from the
Agencia Nacional de Investigaci\'on y Desarrollo (ANID) via Proyecto Fondecyt Regular 1191366; and from ANID BASAL projects CATA-Puente ACE210002 and CATA2-FB210003.  This research has made use of the VizieR catalogue access tool. 
\end{acknowledgements}

\bibliographystyle{aa}
\bibliography{biblio}{}

\end{document}